\documentclass[11pt]{article}

\usepackage[tmargin=3cm,bmargin=3cm]{geometry}

\usepackage[colorlinks]{hyperref}
\hypersetup{
    colorlinks=false,
    pdfborder={0 0 0},
}

\usepackage[round]{natbib}
\renewcommand{\cite}{\citep}

\usepackage[ruled]{algorithm2e}
\renewcommand{\algorithmcfname}{ALGORITHM}
\SetAlFnt{\small}
\SetAlCapFnt{\small}
\SetAlCapNameFnt{\small}
\SetAlCapHSkip{0pt}
\IncMargin{-\parindent}

\usepackage{graphicx}
\usepackage{amsmath}
\usepackage{amssymb}
\usepackage{array}
\usepackage{mathtools}

\DeclareMathOperator*{\argmax}{\mathrm{argmax}}
\DeclareMathOperator*{\argmin}{\mathrm{argmin}}
\newcommand{\eq}[1]{(eq.~\ref{#1})}

\makeatletter
\renewcommand*\env@matrix[1][c]{\hskip -\arraycolsep
  \let\@ifnextchar\new@ifnextchar
  \array{*\c@MaxMatrixCols #1}}
\makeatother
\newcommand{\inner}[1]{\left\langle #1 \right\rangle}

\newcommand{\xhdr}[1]{\vspace{0mm}\noindent{{\bf #1.}}}

\begin{document}

\title{Discovering Social Circles in Ego Networks}
\author{Julian McAuley and Jure Leskovec \\ Stanford\\ \url{jmcauley@cs.stanford.edu}, \url{jure@cs.stanford.edu}}

\maketitle

\begin{abstract}
People's personal social networks are big and cluttered, and currently there is no good way to automatically organize them. Social networking sites allow users to manually categorize their friends into {\em social circles} (e.g.~`circles' on Google+, and `lists' on Facebook and Twitter), however they are laborious to construct and must be updated whenever a user's network grows. In this paper, we study the novel task of automatically identifying users' social circles. We pose this task as a multi-membership node clustering problem on a user's ego-network, a network of connections between her friends. We develop a model for detecting circles that combines network structure as well as user profile information. For each circle we learn its members and the circle-specific user profile similarity metric. Modeling node membership to multiple circles allows us to detect overlapping as well as hierarchically nested circles. Experiments show that our model accurately identifies circles on a diverse set of data from Facebook, Google+, and Twitter, for all of which we obtain hand-labeled ground-truth.
\end{abstract}

\section{Introduction}

Online social networks allow us to follow streams of posts generated by hundreds of our friends and acquaintances.
The people we follow generate overwhelming volumes of information  and to cope with the `information overload' we need to organize our personal social networks \citep{agarwal,harr,khalid}.
One of the main mechanisms for users of social networking sites to organize their networks and the content generated by them is to categorize their friends into what we refer to as \emph{social circles}. Practically all major social networks provide such functionality, for example, `circles' on Google+, and `lists' on Facebook and Twitter. Once a user creates her circles, they  can be used for content filtering,
for privacy,
and for sharing groups of users that others may wish to follow.

\begin{figure}[t]
 \hspace{7mm}\includegraphics[width=0.87\textwidth]{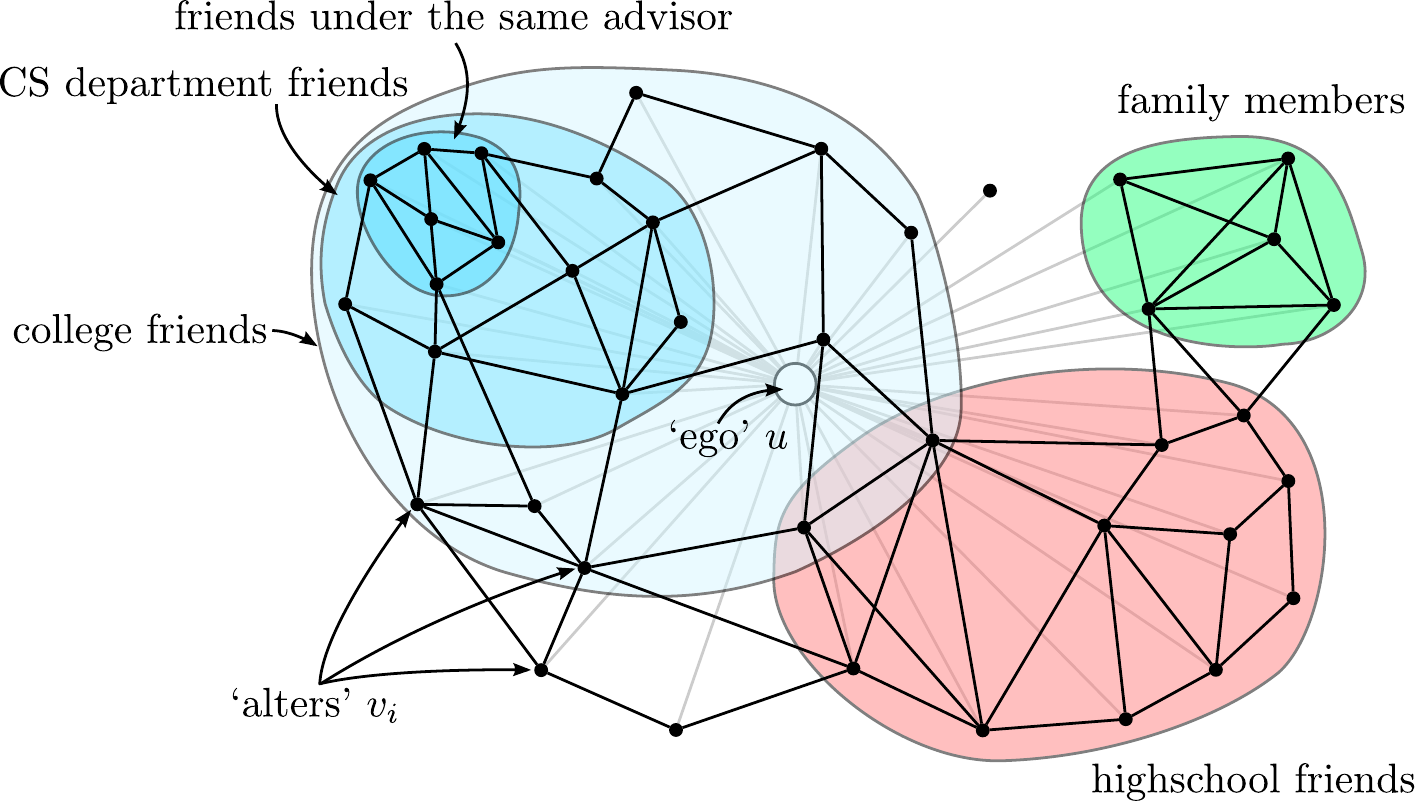}
\caption{An ego-network with labeled circles. The central user, the `ego', is friends with all other users (the `alters') in the network. Alters may belong to any number of circles, including none. We aim to discover circle memberships and to find common properties around which circles form.
This network shows typical behavior that we observe in our data: Approximately 25\% of our ground-truth circles (from Facebook) are contained \emph{completely} within another circle, 50\% overlap with another circle, and 25\% of the circles have no members in common with any other circle.
}
\label{fig:ego}
\end{figure}

Examples of circles from a user's personal social network are shown in Figure \ref{fig:ego}. The `owner' of such a network (the `ego') may form circles based on common bonds and attributes between themselves and the users whom they follow. In this example, the ego may wish to share their latest TKDD article only with their friends from the computer science department, while their baby photos should be shared only with their immediate family; similarly, they may wish to limit the amount of content generated by their high-school friends. These are precisely the types of functionality that \emph{circles} are intended to facilitate.

Currently, users in Facebook, Google+ and Twitter identify their circles either manually, or in a na\"ive fashion by identifying friends sharing a common feature.  Neither approach is particularly satisfactory: the former is time consuming and does not update automatically as a user adds more friends, while the latter fails to capture individual aspects of users' communities, and may function poorly when profile information is missing or withheld.

In this paper we study the problem of automatically discovering users' social circles. In particular, given a single user with her personal social network, our goal is to identify her circles, each of which is a subset of her friends.

Circles are user-specific as each user organizes her personal network of friends independently of all other users to whom she is not connected. This means that we can formulate the problem of circle detection as a clustering problem on her ego-network, the network of friendships between her friends. In practice, circles may overlap (a circle of friends from the same hometown may overlap with a circle from the same college), or be hierarchically nested (among friends from the same college there may be a denser circle from the same degree program). We design our model with both types of behavior in mind.

In Figure~\ref{fig:ego} we are given a single user $u$ and we form a network between her friends $v_i$. We refer to the user $u$ as the {\em ego} and to the nodes $v_i$ as {\em alters}. The task is then to identify the circles to which each alter $v_i$ belongs, as in Figure~\ref{fig:ego}. In other words, the goal is to find communities/clusters in $u$'s ego-network.

Generally, there are two useful sources of data that help with this task. The first is the set of edges of the ego-network. We expect that circles are formed by densely-connected sets of alters \citep{modularity}. However, different circles overlap heavily, i.e., alters belong to multiple circles simultaneously \citep{ahn,palla}, and many circles are hierarchically nested in larger ones (as in Figure~\ref{fig:ego}). Thus it is important to model an alter's memberships to multiple circles. Secondly, we expect that each circle is not only densely connected but that its members also share common properties or traits \citep{mislove2010}. Thus we need to explicitly model the different dimensions of user profiles along which each circle emerges.

We model circle affiliations as latent variables, and similarity between alters as a function of common profile information. We propose an {\em unsupervised} method to learn which dimensions of profile similarity lead to densely linked circles.
After developing a model for this problem, we then study the related problems of \emph{updating} a user's circles once new friends are added to the network, and using weak supervision from the user in the form of `seed nodes' to improve classification. For the former problem, we show that given an already-defined set of a users' circles, we can accurately predict to which circles a new user should be assigned. For the latter problem, we show that classification accuracy improves for each seed node that a user provides, though substantial improvements in accuracy are already obtained even with 2-3 seeds.

Our model has two innovations: First, in contrast to mixed-membership models~\citep{airoldi} we predict \emph{hard} assignment of a node to \emph{multiple} circles, which proves critical for good performance \cite{gregory}. Second, by proposing a parameterized definition of profile similarity, we learn the dimensions of similarity along which links emerge \citep{feld86focused,simmel64affiliations}. This extends the notion of homophily~\citep{homophily1,mcpherson01homophily} by allowing different circles to form along different social dimensions, an idea related to the concept of Blau spaces \citep{blau}. We achieve this by allowing each circle to have a different definition of profile similarity, so that one circle might form around friends from the same school, and another around friends from the same location. We learn the model by simultaneously choosing node circle memberships and profile similarity functions so as to best explain the observed data.

We introduce a dataset of 1,143 ego-networks from Facebook, Google+, and Twitter, for which we obtain hand-labeled ground-truth from 5,636 circles.
Experimental results show that by simultaneously considering social network structure as well as user profile information our method performs significantly better than natural alternatives and the current state-of-the-art. Besides being more accurate our method also allows us to generate automatic explanations of why certain nodes belong to common communities. Our method is completely unsupervised, and is able to automatically determine both the number of circles as well as the circles themselves. We show that the same model can be adapted to deal with weak supervision, and to update already-complete circles as new users arrive.

A preliminary version of this article appeared in \citet{nips2012}.

\subsection{Further Related Work}

Although a `circle' is not precisely the same as a `community', our work broadly falls under the umbrella of community detection \citep{LF09_TR,Schaeffer07_survey,jure10community,POM09_TR,newman2004_detect}. While `classical' clustering algorithms assume disjoint communities \citep{Schaeffer07_survey}, many authors have made the observation that communities in real-world networks may overlap \citep{lancichinetti09ovlpbenchmark,Gregory10LabelOverlap,Lancichinetti09overlapping,jaewon2012}, or have hierarchical structure \citep{ravasz03_hierarchical}.

Topic-modeling techniques have been used to uncover `mixed-memberships' of nodes to multiple groups, and extensions allow entities to be attributed with text information. \citet{airoldi} modeled node attributes as latent variables drawn from a Dirichlet distribution, so that each attribute can be thought of as a partial membership to a community. Other authors extended this idea to allow for side-information associated with the nodes and edges \citep{blockLDA,chang2009relational,liu}. A related line of work by \citet{Hoff02latentspace} also used latent node attributes to model edge formation between `similar' users, which they adapted to clustering problems in \citet{handcock07} and \citet{hoff09}.

Classical clustering algorithms tend to identify communities based on node features \citep{hierarchical} or graph structure \citep{ahn,palla}, but rarely use both in concert. Our work is related to \citet{yoshida} in the sense that it performs clustering on social-network data, and \citet{macbd}, which models memberships to multiple communities. Another work closely related to ours is \citet{jaewon2012}, which explicitly models hard memberships of nodes to multiple overlapping communities, though it does so purely based on network information rather than node features. Our inference procedure is also similar to that of \citet{hastings06_inference}, which treats nodes' assignments to communities as a \emph{Maximum a Posteriori} inference problem between a set of interdependent variables.

Finally, \citet{nubbi,menon,menonICDM} and \citet{dvu} model network data similar to ours; like our own work, they model the probability that two nodes will form an edge, though the underlying models do not form communities, so they are not immediately applicable to the problem of circle detection.

The rest of this paper is organized as follows. We propose a generative model for the formation of edges within communities in Section \ref{sec:model}. In Section \ref{sec:learning} we derive an efficient model parameter learning strategy. In Section \ref{sec:extensions} we describe extensions to our model that allow it to be used in semi-supervised settings, in order to help users update and maintain their circles. We describe the datasets that we construct in Section \ref{sec:data}. We give two schemes for automatically constructing parameterized user similarity function from profile data in Section \ref{sec:features}. In Section \ref{sec:fast} we show how to scale the model to large ego-networks. Finally in Section \ref{sec:experiments} we describe our evaluation and experimental results.

\section{A Generative Model for Friendships in Social Circles}
\label{sec:model}

We desire a model of circle formation with the following properties:
\begin{enumerate}
 \item Nodes within circles should have common properties, or `aspects'.
 \item Different circles should be formed by different aspects, e.g.~one circle might be formed by family members, and another by students who attended the same university.
 \item Circles should be allowed to overlap, and `stronger' circles should be allowed to form within `weaker' ones, e.g.~a circle of friends from the same degree program may form within a circle from the same university, as in Figure \ref{fig:ego}.
 \item We would like to leverage both profile information and network structure in order to identify circles.
 \item Ideally we would like to be able to pinpoint \emph{which} aspects of a profile caused a circle to form, so that the model is interpretable by the user.
\end{enumerate}

The input to our model is an ego-network $G = (V, E)$, along with `profiles' for each user $v \in V$. The `center' node $u$ of the ego-network (the `ego') is not included in $G$, but rather $G$ consists only of $u$'s friends (the `alters'). We define the ego-network in this way precisely because creators of circles do not themselves appear in their own circles.  For each ego-network, our goal is to predict a set of circles
$\mathcal C = \lbrace C_1 \ldots C_K \rbrace$, $C_k \subseteq V$, and associated parameter vectors $\theta_k$
that encode how each circle emerged. We encode `user profiles' into pairwise features $\phi(x,y)$ that in some way capture what properties the users $x$ and $y$ have in common. We first describe our model, which can be applied using arbitrary feature vectors $\phi(x,y)$, and in Section \ref{sec:features} we develop several ways to construct feature vectors $\phi(x,y)$ that are suited to our particular application.

We describe a model of social circles that treats circle memberships as latent variables. Nodes within a common circle are given an opportunity to form an edge, which naturally leads to hierarchical and overlapping circles. We will then devise an \emph{unsupervised} algorithm to jointly optimize the latent variables and the profile similarity parameters so as to best explain the observed network data.

Our model of social circles is defined as follows. Given an ego-network $G$ and a set of $K$ circles $\mathcal C = \lbrace C_1 \ldots C_K \rbrace$, we model the probability that a pair of nodes $(x,y) \in V \times V$ form an edge as
\begin{equation}
 p((x,y) \in E) \propto \exp\Biggl\lbrace\underbrace{\vphantom{\sum_{C_k \nsupseteq \lbrace x,y \rbrace}\alpha_k \inner{\phi(x,y), \theta_k}}\sum_{C_k \supseteq \lbrace x,y \rbrace} \inner{\phi(x,y), \theta_k}}_{\text{circles containing both nodes}} - \underbrace{\sum_{C_k \nsupseteq \lbrace x,y \rbrace}\alpha_k \inner{\phi(x,y), \theta_k}}_{\text{all other circles}}\Biggr\rbrace.
\label{eq:probedge}
\end{equation}
For each circle $C_k$, $\theta_k$ is the profile similarity parameter that we will learn. The idea is that $\inner{\phi(x,y), \theta_k}$ is high if both nodes belong to $C_k$, and low if either of them do not. The parameter $\alpha_k$ trades-off these two effects, i.e., it trades-off the influence of edges within $C_k$ compared to edges outside of (or crossing) $C_k$.
Since the feature vector $\phi(x,y)$ encodes the similarity between the profiles of two users $x$ and $y$, the parameter vector $\theta_k$ encodes which dimensions of profile similarity caused the circle to form, so that nodes within a circle $C_k$ should `look similar' according to $\theta_k$.
Note that the pair $(x,y)$ should be treated as an \emph{unordered} pair in the case of an undirected network (e.g.~Facebook), but should be treated as an \emph{ordered} pair for directed networks (e.g.~Google+ and Twitter).

Considering that edges $e=(x,y)$ are generated independently, we can write the probability of $G$ as
\begin{equation}
 P_\Theta(G; \mathcal C) = \prod_{e \in E} p(e \in E) \times \prod_{e \not\in E} p(e \notin E),
\end{equation}
where $\Theta = \lbrace(\theta_k,\alpha_k)\rbrace^{k=1\ldots K}$ is our set of model parameters. Defining the shorthand notation
\begin{equation*}
d_k(e) = \delta(e \in C_k) - \alpha_k \delta(e \notin C_k),\quad
\Phi(e) = \sum_{C_k \in \mathcal C} d_k(e) \inner{\phi(e), \theta_k}
\end{equation*}
allows us to write the log-likelihood of $G$:
\begin{equation}
 l_\Theta(G; \mathcal C) = \sum_{e \in E} \Phi(e) - \sum_{e \in V\times V}\log\left(1 + e^{\Phi(e)}\right),
\label{eq:ll}
\end{equation}
where $Z = (1 + e^{\Phi(e)})$ is a normalization constant.

Next, we describe how to optimize node circle memberships $\mathcal C$ as well as the parameters of the user profile similarity functions $\Theta=\lbrace(\theta_k,\alpha_k)\rbrace$ ($k=1\ldots K$) given a graph $G$ and user profiles.

\section{Unsupervised Learning of Model Parameters}
\label{sec:learning}

Treating circles ${\mathcal C}$ as latent variables, we aim to find $\hat{\Theta}=\{\hat{\theta}, \hat{\alpha}\}$ so as to maximize the regularized log-likelihood of \eq{eq:ll}, i.e.,
\begin{equation}
 \hat{\Theta}, \hat{\mathcal C} = \argmax_{\Theta, {\mathcal C}} l_\Theta(G; \mathcal C) - \lambda\Omega(\theta).
\label{eq:argmax}
\end{equation}
We solve this problem using coordinate ascent on $\Theta$ and $\mathcal C$ \cite{MacKay}:
\begin{eqnarray}
 \mathcal C^t & = & \argmax_{\mathcal C} l_{\Theta^t}(G; \mathcal C) \label{eq:opt1}\\
 \Theta^{t+1} & = & \argmax_{\Theta} l_\Theta(G; \mathcal C^t) - \lambda \Omega(\theta). \label{eq:opt2}
\end{eqnarray}
We optimize \eq{eq:opt2} using L-BFGS, a standard quasi-Newton procedure to optimize smooth functions of many variables \citep{lbfgs}. Partial derivatives are given by
\begin{eqnarray}
 \frac{\partial l}{\partial \theta_k} &=& \sum_{e \in V \times V} -d_e(k) \phi(e)_k \frac{e^{\Phi(e)}}{1 + e^{\Phi(e)}} +  \sum_{e \in E} d_k(e)\phi(e)_k - \frac{\partial\Omega}{\partial \theta_k}\label{eq:partial1}\\
 \frac{\partial l}{\partial \alpha_k} &=& \sum_{e \in V \times V} \delta(e \notin C_k)\inner{\phi(e), \theta_k}\frac{e^{\Phi(e)}}{1 + e^{\Phi(e)}} - \sum_{e \in E} \delta(e \notin C_k)\inner{\phi(e), \theta_k}.\label{eq:partial2}
\end{eqnarray}

For fixed $\mathcal C \setminus C_i$ we note that solving $\argmax_{C_i} l_\Theta (G; \mathcal C \setminus C_i)$ can be expressed as pseudo-boolean optimization in a pairwise graphical model \citep{boros}. `Pseudo-boolean optimization' refers to problems defined over boolean variables (in this case, whether or not a node is assigned to a particular community), where the variables being optimized are interdependent (in this case, relationships are defined over edges in a graph). In short, our optimization problem can be written in the form
\begin{equation}
 C_k = \argmax_C \!\!\sum_{(x,y) \in V\times V}\!\! E^k_{(x,y)}(\delta(x \in C), \delta(y \in C)).
\label{eq:pseudo}
\end{equation}
Although this problem class is NP-hard in general, efficient approximation algorithms are readily available \citep{rother07}.
In our setting, we want edges with high weight (under $\theta_k$) to appear in $C_k$, and edges with low weight to appear outside of $C_k$. Defining $$o_k(e) = \sum_{C_k \in \mathcal C \setminus C_i} d_k(e)\inner{\phi(e), \theta_k}$$ the energy $E^k_e$ of \eq{eq:pseudo} is
\small
\begin{eqnarray*}
E^k_e(0,0) = E^k_e(0,1) = E^k_e(1,0) & = & \left\lbrace \begin{array}{ll} o_k(e) - \alpha_k\inner{\phi(e), \theta_k} - \log(1 + e^{o_k(e) - \alpha_k\inner{\phi(e), \theta_k}}), & e \in E\\
                                  -\log(1 + e^{o_k(e) - \alpha_k\inner{\phi(e), \theta_k}}), & e \notin E
                                 \end{array}\right.\\
E^k_e(1,1) & = & \left\lbrace \begin{array}{ll} o_k(e) + \inner{\phi(e), \theta_k} - \log(1 + e^{o_k(e) + \inner{\phi(e), \theta_k}}), & e\in E\\
                             -\log(1 + e^{o_k(e) + \inner{\phi(e), \theta_k}}), & e \notin E
                            \end{array}\right..
\end{eqnarray*}
\normalsize
By expressing the problem in this form we can draw upon existing work on pseudo-boolean optimization. We use the publicly-available `QPBO' software described in \citet{rother07}, which implements algorithms described in \citet{hammer84} and \citet{kohli05}, and is able to accurately approximate problems of the form shown in \eq{eq:pseudo}. Essentially, problems of the type shown in \eq{eq:pseudo} are reduced to \emph{maximum flow}, where boolean labels for each node are recovered from their assignments to `source' and `sink' sets. Such algorithms have worst-case complexity $O(|N|^3)$, though the average case running-time is far better \citep{kolmo07}. We solve \eq{eq:pseudo} for each $C_k$ in a random order.

The two optimization steps of \eq{eq:opt1} and \eq{eq:opt2} are repeated until convergence, i.e., until $\mathcal C^{t+1} = \mathcal C^t$. The entire procedure is presented in Algorithm \ref{alg:vanilla}. We regularize \eq{eq:argmax} using the $\ell_1$ norm, i.e., $$\Omega(\theta) = \sum_{k=1}^K\sum_{i=1}^{|\theta_k|} |\theta_{ki}|,$$ which leads to sparse (and readily interpretable) parameters. Our algorithm can readily handle all but the largest problem sizes typically observed in ego-networks: in the case of Facebook, the average ego-network has around 190 nodes \citep{nfriends}, while the largest network we encountered has 4,964 nodes. Later, in Section \ref{sec:fast}, we will exploit the fact that our features are binary, and that many nodes share similar features, to develop more efficient algorithms based on Markov Chain Monte Carlo inference.
Note that since the method is \emph{unsupervised}, inference is performed independently for each ego-network. This means that our method could be run on the full Facebook graph (for example), as circles are independently detected for each user, and the ego-networks typically contain only hundreds of nodes. In Section \ref{sec:extensions} we describe extensions that allow our model to be used in semi-supervised settings.

\begin{algorithm}
\label{alg:vanilla}
 \SetAlgoLined
 \KwData{ego-network $G = (V,E)$, edge features $\phi(e) : E \rightarrow \mathbb R^F$, hyperparameters $\lambda$, $K$}
 \KwResult{parameters $\hat\Theta \coloneqq \lbrace (\hat{\theta}_k, \hat{\alpha}_k) \rbrace^{k = 1 \ldots K}$, communities $\hat{\mathcal C}$}
 initialize $\theta^0_k \in \lbrace 0, 1 \rbrace^F$, $\alpha^0_k \coloneqq 1$, $C_k \coloneqq \varnothing$, $t \coloneqq 0$\;
 \Repeat{$\mathcal \mathcal C^{t + 1} = \mathcal C^t$}{
   \For{$k \in \lbrace 1 \ldots K \rbrace$}{
     $\mathcal C_k^t \coloneqq \argmax_C \!\!\sum_{(x,y) \in V\times V}\!\! E^k_{(x,y)}(\delta(x \in C), \delta(y \in C))$\; \tcp{using QPBO, see \eq{eq:pseudo}}
   }
   $\Theta^{t+1} \coloneqq \argmax_{\Theta} l_\Theta(G; \mathcal C^t) - \lambda \Omega(\theta)$\; \tcp{using L-BFGS, see (eqs.~\ref{eq:partial1} and \ref{eq:partial2})}
   $t \coloneqq t + 1$\;
 }
 \caption{Predict complete circles with hyperparameters $\lambda$, $K$.}
\end{algorithm}

\subsection{Hyperparameter Estimation}
\label{sec:hyper}

To choose the optimal number of circles, we choose $K$ so as to minimize an approximation to the Bayesian Information Criterion (BIC), an idea seen in several works on probabilistic clustering \citep{airoldi,handcock,volinsky}. In this context, the Bayesian Information Criterion is defined as
\begin{equation}
 \mathit{BIC}(K;\Theta^K) \simeq -2 l_{\Theta^K}(G; \mathcal C) + |\Theta^K|\log |E|,
\end{equation}
where $\Theta^K$ is the set of parameters predicted when there are $K$ circles, and $|\Theta^K|$ is the number of parameters (which increases linearly as $K$ increases). We then choose $K$ so as to minimize this objective:
\begin{equation}
 \hat{K} = \argmin_K \mathit{BIC}(K;\Theta^K).
\label{eq:barK}
\end{equation}
In other words, an additional circle will only be added to the model if doing so has a `significant' impact on the log-likelihood.

The regularization parameter $\lambda \in \lbrace 0, 1, 10, 100 \rbrace$ was determined using leave-one-out cross validation, though in our experience did not significantly impact performance.

\section{Extensions}
\label{sec:extensions}

So far, we have considered the `cold-start' problem of predicting complete sets of circles using nothing but node attributes and edge information. In other words, we have treated circle prediction as an \emph{unsupervised} task.
This setting is realistic if users construct their circles only after their ego-networks have already been defined.
On the other hand, in settings where users build their circles incrementally, it is less likely that we would wish to predict complete circles `from scratch'. We note that both settings occur in the three social networks that we consider.

In this section, we describe techniques to exploit partially observed circle information to help users update and maintain their circles. In other words, we would like to apply our model to users' personal networks as they change and evolve \citep{lars06groups}. Since our model is probabilistic, it is straightforward to adapt it to make use of partially observed data, by conditioning on the assignments of some of the latent variables in our model. In this way, we adapt our model for semi-supervised settings in which a user labels some or all of the members of their circles. Later, in Section \ref{sec:fast}, we describe modifications of our model that allow it to be applied to extremely large networks, by exploiting the fact that many users assigned to common circles also have common features.

\subsection{Circle Maintenance}
\label{sec:newnode}

First we deal with the problem of a user adding new friends to an established ego-network, whose circles have already been defined. Thus, given a complete set of circles, our goal is to predict community memberships for a new node, based on that node's features, and their patterns of connectivity to existing nodes in the ego-network.

Since circles in this setting are fully-observed, we simply fit the model parameters that best explain the ground-truth circles $\bar{\mathcal C}$ provided by the user:
\begin{equation}
\hat{\Theta} = \argmax_{\Theta} l_\Theta(G; \bar{\mathcal C}) - \lambda \Omega(\theta). \label{eq:opt3}
\end{equation}
As with \eq{eq:opt2} this is solved using L-BFGS, though optimization is significantly faster in this case as there are no longer latent community memberships to infer, and thus coordinate ascent is not required.

Next, we must predict to which of the $K$ ground-truth circles a new user $u$ belongs. That is, we must predict $c^u \in \lbrace 0, 1 \rbrace^K$, where each $c^u_k$ is a binary variable indicating whether the user $u$ should belong to the circle $C_k$. In practice, for the sake of evaluation, we shall suppress a single user from $G$ and $\bar{\mathcal C}$, and try to recover their memberships.

This can be done by choosing the assignment $c^u$ that maximizes the log-likelihood of $\mathcal C$ once $u$ is added to the graph. We define the augmented community memberships as $\mathcal C^+ = \lbrace C_k^+ \rbrace^{k = 1 \ldots K}$, where
\begin{equation}
C_k^+ = \left\lbrace \begin{array}{ll}
                      \bar{C}_k \cup \lbrace u \rbrace, & c^u_k = 1\\
                      \bar{C}_k, & c^u_k = 0
                     \end{array}
 \right..
\end{equation}
The updated community memberships are then chosen according to
\begin{equation}
\hat{\mathcal C}^+ = \argmax_{c^u} l_{\hat{\Theta}}(G \cup \lbrace u \rbrace; \mathcal C^+) \label{eq:opt4}.
\end{equation}
The above expression can be computed efficiently for different values of $c^u$ by noting that the log-likelihood only changes for terms including $u$, meaning that we need to compute $p((x,y) \in E)$ only if $x = u$ or $y = u$. In other words, we only need to consider how the new user relates to existing users, rather than considering how existing users relate to each other; thus computing the log-likelihood requires linear (rather than quadratic) time. To find the optimal $c^u$ we can simply enumerate all $2^K$ possibilities, which is feasible so long as the user has no more than $K \simeq 20$ circles. For users with more circles we must resort to an iterative update scheme as we did in Section \ref{sec:learning}.

\subsection{Semi-Supervised Circle Prediction}
\label{sec:seednodes}

Next, we consider the problem of using weak supervision in the form of `seed nodes' to assist in circle prediction \citep{andersen06seed}. In this setting, the user manually labels a few users from each of the circles they want to create, say $\lbrace s_1 \ldots s_K \rbrace$. Our goal is then to predict $K$ circles $\mathcal C = \lbrace C_1 \ldots C_K \rbrace$ subject to the constraint that $s_k \subseteq C_k$ for all $k \in \lbrace 1 \ldots K \rbrace$.

Again, since our model is probabilistic, this can be done by conditioning on the assignments of some of the latent variables. That is, we simply optimize $l_\Theta(G; \mathcal C)$ subject to the constraint that $s_k \subseteq C_k$ for all $k \in \lbrace 1 \ldots K \rbrace$. In the parlance of graphical models, this means that rather than treating the seed nodes as latent variables to be predicted, we treat them as evidence on which we condition. We could also include negative evidence (i.e., the user could provide labels for users who do \emph{not} belong to each circle), or we could have users provide additional labels interactively, though the setting described is the most similar to what is used in practice.

\section{Dataset Description}
\label{sec:data}

Our goal is to evaluate our method on ground-truth data. We expended significant time, effort, and resources to obtain high quality hand-labeled data, which we have made available online.\footnote{\url{http://snap.stanford.edu/data/}} We were able to obtain ego-networks and ground-truth from three major social networking sites: Facebook, Google+, and Twitter.

From Facebook we obtained profile and network data from 10 ego-networks, consisting of 193 circles and 4,039 users. To obtain circle information we developed our own Facebook application and conducted a survey of ten users, who were asked to manually identify all of the circles to which their friends belonged. It took each user between 2 and 3 hours to label their entire network. On average, users identified 19 circles in their ego-networks, with an average circle size of 22 friends. Examples of circles we obtained include students of common universities and classes, sports teams, relatives, etc.

\begin{figure}
\begin{center}
 \includegraphics[scale=0.75]{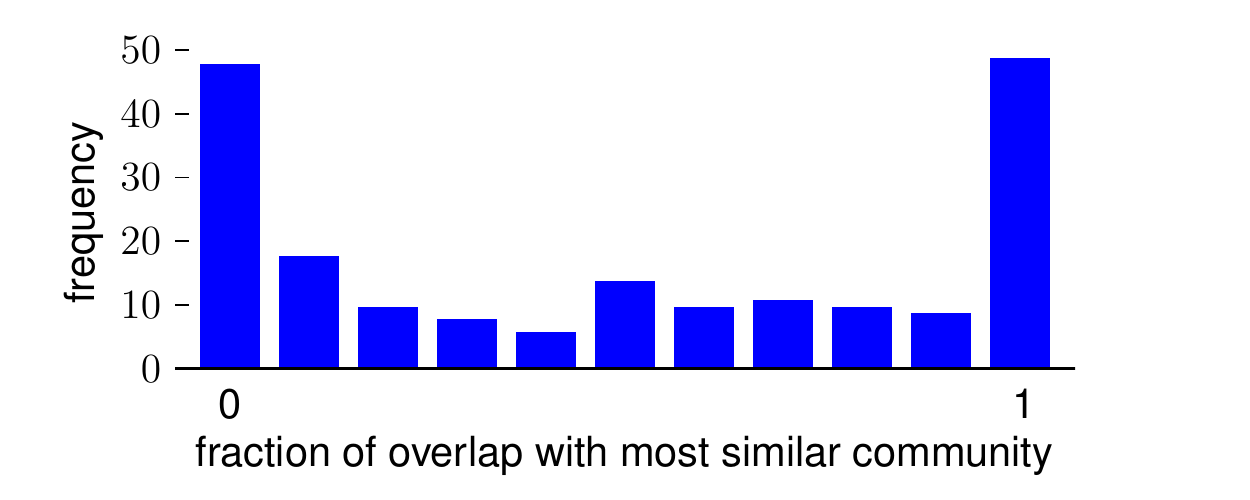}
\end{center}
\caption{Histogram of overlap between circles (on Facebook). A value of zero indicates that the circle does not intersect with any of the user's other circles, whereas a value of one indicates that a circle is entirely contained within another. Approximately $25\%$ of circles exhibit the latter behavior.}
\label{fig:overlap}
\end{figure}

Figure \ref{fig:overlap} shows the extent to which our 193 user-labeled circles in 10 ego networks from Facebook overlap (intersect) with each other. Around one quarter of the identified circles are independent of any other circle, though a similar fraction are completely contained within another circle (e.g.~friends who studied under the same adviser may be a subset of friends from the same university). The remaining $50\%$ of communities overlap to some extent with another circle.

For the other two datasets we obtained publicly accessible data. From Google+ we obtained data from 133 ego-networks, consisting of 479 circles and 106,674 users. The 133 ego-networks represent all 133 Google+ users who had shared at least two circles, and whose network information was publicly accessible at the time of our crawl. The Google+ circles are quite different to those from Facebook, in the sense that their creators have chosen to release them publicly, and because Google+ is a \emph{directed} network (note that our model can very naturally be applied to both to directed and undirected networks). For example, one circle contains candidates from the 2012 republican primary, who presumably do not follow their followers, nor each other.
Finally, from Twitter we obtained data from 1,000 ego-networks, consisting of 4,869 circles (or `lists' \citep{lists3,tadvise,twitterLists,twitterLists2}) and 81,362 users. The ego-networks we obtained range in size from 10 to 4,964 nodes.

Taken together our data contains 1,143 different ego-networks, 5,541 circles, and 192,075 users. The size differences between these datasets simply reflects the availability of data from each of the three sources. Our Facebook data is \emph{fully labeled}, in the sense that we obtain \emph{every} circle that a user considers to be a cohesive community, whereas our Google+ and Twitter data is only \emph{partially labeled}, in the sense that we only have access to public circles. We design our evaluation procedure in Section \ref{sec:experiments} so that partial labels cause no issues.

\section{Constructing Features from User Profiles}
\label{sec:features}

\begin{figure*}[t]
\begin{center}
\includegraphics[width=0.75\textwidth]{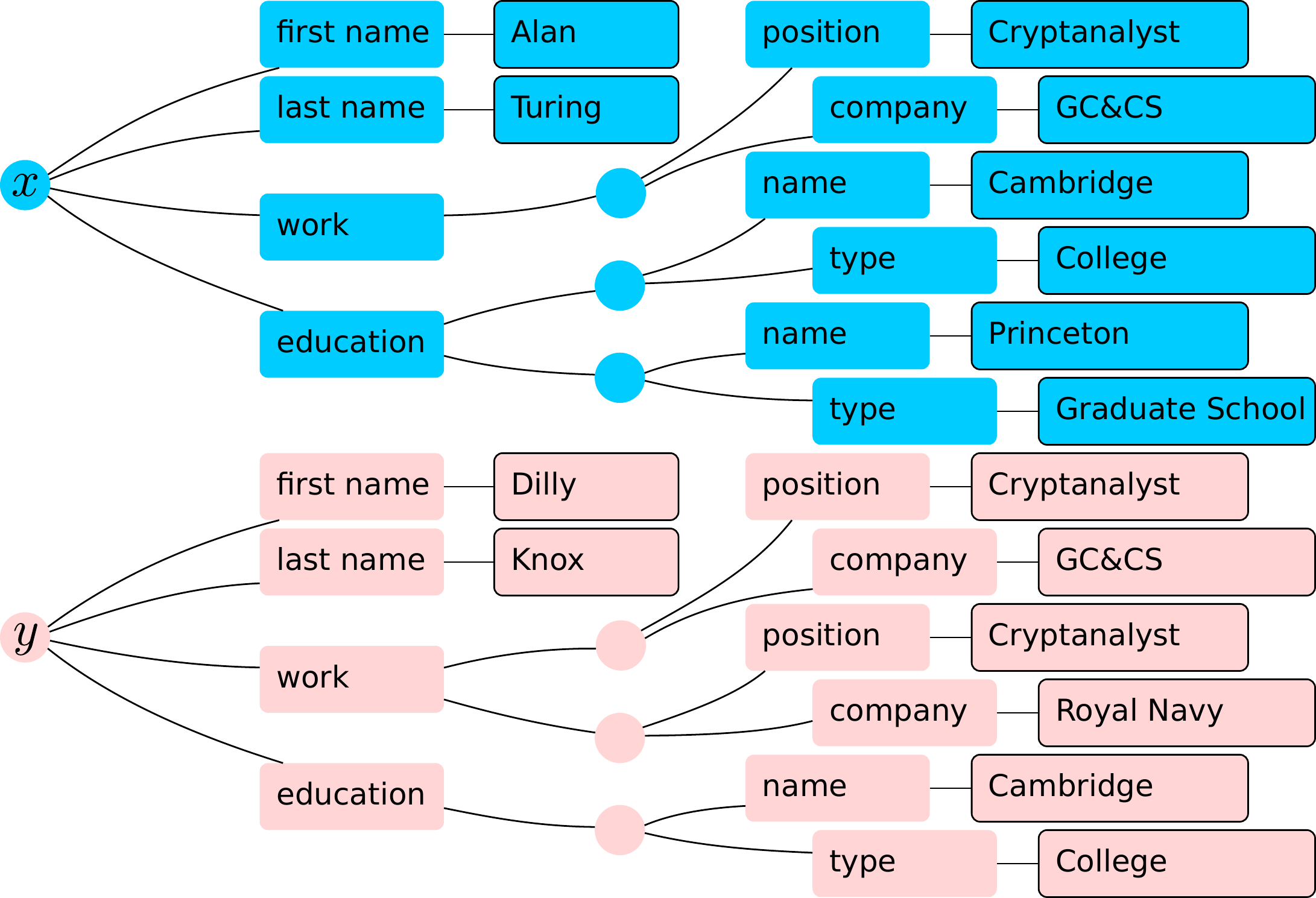}

\vspace{5mm}
    \parbox[c]{0.59\textwidth}{\centering
        $\mathbf{1} - \sigma_{x,y} = \left[\begin{matrix} 0 \\ 0 \\ 0 \\ 0 \\ 1 \\ 1 \\ 0 \\ 1 \\ 1 \\ 0 \\ 0 \end{matrix}\right]  \begin{matrix}[l] \mathit{first\ name:Dilly} \\ \mathit{last\ name:Knox} \\ \mathit{first\ name:Alan} \\ \mathit{last\ name:Turing} \\ \mathit{work:position:Cryptanalyst} \\ \mathit{work:location:GC\&CS} \\ \mathit{work:location:Royal\  Navy} \\ \mathit{education:name:Cambridge} \\ \mathit{education:type:College} \\ \mathit{education:name:Princeton} \\ \mathit{education:type:Graduate\ School} \end{matrix}$}
\parbox[c]{0.35\textwidth}{\centering
        $\mathbf{1} - \sigma'_{x,y} = \left[\begin{matrix} 0 \\ 0 \\ 1 \\ 1 \\ 1 \\ 1 \end{matrix}\right]  \begin{matrix}[l] \mathit{first\ name} \\ \mathit{last\ name} \\ \mathit{work:position} \\ \mathit{work:location} \\ \mathit{education:name} \\ \mathit{education:type}\end{matrix}$}
\end{center}
\caption{Feature construction. Profiles are tree-structured, and we construct features by comparing paths in those trees. Examples of trees for two users $x$ (blue) and $y$ (pink) are shown at top. Two schemes for constructing feature vectors from these profiles are shown at bottom: (1) (bottom left) we construct binary indicators measuring the difference between leaves in the two trees, e.g. `work$\rightarrow$position$\rightarrow$Cryptanalyst' appears in both trees. (2) (bottom right) we sum over the leaf nodes in the first scheme, maintaining the fact that the two users worked at the same institution, but discarding the \emph{identity} of that institution.}
\label{fig:trees}
\end{figure*}

Profile information in all of our datasets can be represented as a \emph{tree} where each level encodes increasingly specific information (Figure~\ref{fig:trees}, left). In other words, user profiles are organized into increasingly specific categories. For example, a user's profile might have a \emph{education} category, which would be further separated into categories such as \emph{name}, \emph{location}, and \emph{type}. The leaves of the tree are then specific values in these categories, e.g.~\emph{Princeton}, \emph{Cambridge}, and \emph{Graduate School}. Several works deal with automatically building features from tree-structured data \citep{haussler99convolution,treekernels}, but in order to understand the relationship between circles and user profile information, we shall design our own feature representation scheme.

We propose two hypotheses for how users organize their social circles: either they may form circles around users who share some common property \emph{with each other}, or they may form circles around users who share some common property \emph{with themselves}. For example, if a user has many friends who attended Stanford, then they may form a `Stanford' circle. On the other hand, if they themselves did \emph{not} attend Stanford, they may not consider attendance to Stanford to be a salient feature. The feature construction schemes we propose allow us to assess which of these hypotheses better represents the data we obtain.

From Google+ we collect data from six categories (gender, last name, job titles, institutions, universities, and places lived). From Facebook we collect data from 26 categories, including users' hometowns, birthdays, colleagues, political and religious affiliations, etc. As a proxy for profile data, from Twitter we collect data from two categories, namely the set of hashtags and mentions used by each user during two-weeks' worth of tweets. `Categories' correspond to parents of leaf nodes in a profile tree, as shown in Figure \ref{fig:trees}.

We first propose a difference vector to encode the relationship between two profiles. A non-technical description is given in Figure \ref{fig:trees}. Essentially, we want to encode those dimensions where two users are the same (e.g.~Alan and Dilly went to the same graduate school), and those where they are different (e.g.~they do not have the same surname). Suppose that users $v\in V$ each have an associated profile tree $T_v$, and that $l \in T_v$ is a leaf in that tree. We define the difference vector $\sigma_{x,y}$ between two users $x$ and $y$ as a binary indicator encoding the profile aspects where users $x$ and $y$ differ (Figure \ref{fig:trees}, bottom left):
\begin{equation}
\sigma_{x,y}[l] = \delta((l \in \mathcal T_x) \neq (l \in \mathcal T_y)).
\end{equation}
Note that feature descriptors are defined \emph{per ego-network}: while many thousands of high schools (for example) exist among all Facebook users, only a small number appear among any particular user's friends.

Although the above difference vector has the advantage that it encodes profile information at a fine granularity, it has the disadvantage that it is high-dimensional (up to 4,122 dimensions in the data we considered).
One way to address this is to form difference vectors based on the \emph{parents} of leaf nodes: this way, we encode what profile \emph{categories} two users have in common, but disregard specific values (Figure \ref{fig:trees}, bottom right). For example, we encode \emph{how many} hashtags two users tweeted in common, but discard \emph{which} hashtags they tweeted:
\begin{equation}
\sigma'_{x,y}[p] = \textstyle{\sum_{l \in \mathit{children}(p)}} \sigma_{x,y}[l].
\label{eq:compressed}
\end{equation}
This scheme has the advantage that it requires a {\em constant} number of dimensions, regardless of the size of the ego-network (26 for Facebook, 6 for Google+, 2 for Twitter, as described above).

Based on the difference vectors $\sigma_{x,y}$ (and $\sigma'_{x,y}$) we now describe how to construct edge features $\phi(x,y)$. The first property we wish to model is that \emph{members of circles should have common relationships} with each other:
\begin{equation}
 \phi^1(x,y) = (1; -\sigma_{x,y}).
\label{eq:phi1}
\end{equation}
The second property we wish to model is that \emph{members of circles should have common relationships to the ego of the ego-network}. In this case, we consider the profile tree $T_u$ \emph{from the ego user $u$}.
We then define our features in terms of that user:
\begin{equation}
 \phi^2(x,y) = (1; -\bigl|\sigma_{x,u} - \sigma_{y,u}\bigr|)
\label{eq:phi2}
\end{equation}
($|\sigma_{x,u} - \sigma_{y,u}|$ is taken elementwise).
These two parameterizations allow us to assess which mechanism better captures users' subjective definition of a circle.
In both cases, we include a constant feature (`1'),
which controls the probability that edges form within circles, or equivalently it measures the extent to which circles are made up of friends. Importantly, this allows us to predict memberships even for users who have no profile information, simply due to their patterns of connectivity.

Similarly, for the `compressed' difference vector $\sigma'_{x,y}$, we define
\begin{equation}
  \psi^1(x,y) = (1; -\sigma'_{x,y}), \quad \psi^2(x,y) = (1; -\bigl|\sigma'_{x,u} - \sigma'_{y,u}\bigr|) \label{eq:psi2}.
\end{equation}
To summarize, we have identified four ways of representing the compatibility between different aspects of profiles for two users. We considered two ways of constructing a difference vector ($\sigma_{x,y}$ vs.~$\sigma'_{x,y}$) and two ways of capturing the compatibility between a pair of profiles ($\phi(x,y)$ vs. $\psi(x,y)$). The features are designed to model the following behavior:
\begin{enumerate}
 \item Ego users build circles around common relationships between their friends ($\phi^1$, $\psi^1$)
 \item Ego users build circles around common relationships between their friends and themselves ($\phi^2$, $\psi^2$)
\end{enumerate}
In our experiments we assess which of these assumptions is more realistic in practice.

\section{Fast Inference in Large Ego-Networks}
\label{sec:fast}

Although our algorithm is able to handle the problem sizes typically encountered in ego-networks (i.e., fewer than 1,000 friends), scalability to larger networks presents an issue, as we require quadratic memory to encode the compatibility between every pair of nodes (an issue we note is also present in the existing approaches we consider in Section \ref{sec:experiments}). In this section, we propose a more scalable alternative that makes use of the fact that many nodes belonging to common communities also share common features.

Noting that features $\phi^1$ and $\phi^2$ described in Section \ref{sec:features} are binary valued, as are community memberships, if there are $K$ communities and $F$-dimensional features, there can be at most $2^{K + F}$ `types' of node. In other words, every node's community membership is drawn from $\lbrace 0, 1 \rbrace^K$, and every node's feature vector is drawn from $\lbrace 0, 1 \rbrace^F$, so there are at most $2^{K + F}$ distinct community/feature combinations. Of course the number of distinct node types is also bounded by $|V|$, the number of nodes in the graph.

In practice, however, the number of distinct node `types' is much smaller, as nodes belonging to common communities tend to have common features. Community memberships are also not independent: in Figure \ref{fig:overlap} we observed both disjoint and hierarchically nested communities, which means that of the $2^K$ possible community memberships, only a fraction of them occur in practice.

In this section, we propose a Markov-Chain Monte Carlo (MCMC) sampler \citep{Newman99MonteCarlo} which efficiently updates node-community memberships by `collapsing' nodes that have common features and community memberships. Note that the adaptations to be described can be applied to \emph{any} types of feature (i.e., not just binary features), all we require is that many users share the same features; we assume binary features merely for the sake of presentation.

We start by representing each node using binary strings that encode both its community memberships and its features. Each node's community memberships are represented using $S : V \rightarrow \Sigma^K$, such that
\begin{equation}
 S(x)[k] = \left\lbrace \begin{array}{ll} 1, & \text{if\ }x \in C_k\\ 0, & \text{otherwise} \end{array} \right..
\end{equation}
Similarly, each node's features are represented using the binary string $Q$, which, since our features are already binary, is simply the concatenation of the feature dimensions.

We now say that the `type' of a node $x$ is the concatenation of its community string and its feature string, $(S(x);Q(x))$, and we build a (sparse) table $\mathit{types} : \Sigma^K \times \Sigma^F \rightarrow \mathbb N$ that counts how many nodes exist of each type.

In our setting, MCMC consists of repeatedly updating the (binary) label of each node in a particular community. Specifically, if the marginal (log) probability that a node $x$ belongs to a community $k$ is given by $\ell^k_x$, then the node's new label is chosen by sampling $z \leftarrow \mathcal U(0,1)$, and updating
\begin{equation}
 S(x)[k] = \left\lbrace \begin{array}{ll}
                         1, & \text{if\ } z < \exp{\lbrace \frac{1}{T}(\ell^k_x(1) - \ell^k_x(0)) \rbrace}\\
                         0, & \text{otherwise}
                        \end{array}
\right.,
\end{equation}
where $T$ is a temperature parameter that decreases at each iteration, so that we are more likely to choose the label with higher probability as the model `cools'.

Computing $\ell^k_x(0)$ and $\ell^k_x(1)$ (the probability that node $x$ takes the label $0$ or $1$ in community $k$) requires computing $p((x,y) \in E)$ for all $y \in V$. However, we note that if two nodes $y$ and $y'$ have the same type (i.e., they belong to the same communities and have the same features), then $p((x,y) \in E) = p((x,y') \in E)$. In order to maximize the log-likelihood of the observed data, we must also consider whether $(x,y)$ and $(x,y')$ are actually edges in the graph. To do so, we first compute $\ell^k_x(0)$ and $\ell^k_x(1)$ under the assumption that no edges are incident on $x$, after which we correct for those edges incident on $x$. Thus the running time of a single update is linear in the number of distinct node types, plus the average node degree, both of which are bounded by the number of nodes.

The entire procedure is demonstrated in Algorithm \ref{alg:mcmc}.

\begin{algorithm}
\label{alg:mcmc}
 \SetAlgoLined
 \KwData{node $x$ whose membership to circle $C_k$ is to be updated
}
 \KwResult{updated membership for node $x$}
 initialize $\ell^k_x(0) \coloneqq 0$, $\ell^k_x(1) \coloneqq 0$\;
 construct a dummy node $x_0$ with the communities and features of $x$ but with $x \notin C_k$\;
 construct a dummy node $x_1$ with the communities and features of $x$ but with $x \in C_k$\;
   \For{$(c,f) \in \mathtt{dom}(\mathit{types})$}{
     \tcp{$c = $ community string, $f = $ feature string}
     $n \coloneqq \mathit{types}(c,f)$\;
     \tcp{$n = $ number of nodes of this type}
     \If{$S(x) = c \wedge Q(x) = f$}
     {
       \tcp{avoid including a self-loop on $x$}
       $n \coloneqq n - 1$\;
     }
     construct a dummy node $y$ with community memberships $c$ and features $f$\;
     \tcp{first compute probabilities assuming all pairs $(x,y)$ are non-edges}
     $\ell^k_x(0) \coloneqq \ell^k_x(0) + n \log p((x_0, y) \notin E)$\;
     $\ell^k_x(1) \coloneqq \ell^k_x(1) + n \log p((x_1, y) \notin E)$\;
   }
  \For{$(x,y) \in E$}
  {
    \tcp{correct for edges incident on $x$}
    $\ell^k_x(0) \coloneqq \ell^k_x(0) - \log p((x_0, y) \notin E) + \log p((x_0, y) \in E)$\;
    $\ell^k_x(1) \coloneqq \ell^k_x(1) - \log p((x_1, y) \notin E) + \log p((x_1, y) \in E)$\;
  }
 \tcp{update membership to circle $k$}
 $\mathit{types}(S(x), Q(x)) \coloneqq \mathit{types}(S(x), Q(x)) - 1$\;
 $z \leftarrow \mathcal U(0,1)$\;
 \eIf{$z < \exp{\lbrace T(\ell^k_x(1) - \ell^k_x(0)) \rbrace}$}{$S(x)[k] \coloneqq 1$}{$S(x)[k] \coloneqq 0$}
 $\mathit{types}(S(x), Q(x)) \coloneqq \mathit{types}(S(x), Q(x)) + 1$\;
 \caption{Update memberships node $x$ and circle $k$.}
\end{algorithm}

We also exploit the same observation when computing partial derivatives of the log-likelihood,
that is we first efficiently compute derivatives under the assumption that the graph contains no edges, and then correct the result by summing over all edges in $E$.

\section{Experiments}
\label{sec:experiments}

We first describe the evaluation metrics to be used in Sections \ref{sec:evaluation} and \ref{sec:aligning}, before describing the baselines to be evaluated in Section \ref{sec:baselines}. We describe the performance of our (unsupervised) algorithm in Section \ref{sec:performance}, and extensions in Sections \ref{sec:results_maintenance}, \ref{sec:results_semisupervised}, and \ref{sec:results_scalability}.

\subsection{Evaluation metrics}
\label{sec:evaluation}

Although our method is unsupervised, we can evaluate it on ground-truth data by examining the maximum-likelihood assignments of the latent circles $\mathcal C = \lbrace C_1 \ldots C_K \rbrace$ after convergence. Our goal is that for a properly regularized model, the latent circles will align closely with the human labeled ground-truth circles $\bar{\mathcal C} = \lbrace \bar{C}_1 \ldots \bar{C}_{\bar{K}} \rbrace$.

To measure the alignment between a predicted circle $C$ and a ground-truth circle $\bar{C}$, we compute the Balanced Error Rate (BER) between the two circles \citep{chen06},
\begin{equation}
 \mathit{BER}(C, \bar{C}) = \frac{1}{2}\left( \frac{|C \setminus \bar{C}|}{|C|} + \frac{|\bar{C} \setminus C|}{|\bar{C}|} \right).
\end{equation}
This measure assigns equal importance to false positives and false negatives, so that trivial or random predictions incur an error of $0.5$ on average. Such a measure is preferable to the $0/1$ loss (for example), which assigns extremely low error to trivial predictions. We also report the $F_1$ score, which we find produces qualitatively similar results.

\subsection{Aligning predicted and ground-truth circles}
\label{sec:aligning}
Since we do not know the correspondence between circles in $\mathcal C$ and $\bar{\mathcal C}$, we compute the optimal match via linear assignment by maximizing:
\begin{equation}
 \max_{f : \mathcal C \rightarrow \bar{\mathcal C}} \frac{1}{|f|} {\sum_{C \in \text{dom}(f)}} (1 - \mathit{BER}(C, f(C))),
\label{eq:delta}
\end{equation}
where $f$ is a (partial) correspondence between $\mathcal C$ and $\bar{\mathcal C}$. That is, if the number of predicted circles $|\mathcal C|$ is less than the number of ground-truth circles $|\bar{\mathcal C}|$, then every circle $C \in \mathcal C$ must have a match $\bar{C} \in \bar{\mathcal C}$, but if $|\mathcal C| > |\bar{\mathcal C}|$, we do not incur a penalty for additional predictions that \emph{could} have been circles but were not included in the ground-truth. We use established techniques to estimate the number of circles, so that none of the baselines suffers a disadvantage by mispredicting $\hat{K}=|\mathcal C|$.

In the case of Facebook (where we have `complete' ground-truth, in the sense that survey participants ostensibly label \emph{every} circle), our method ought to penalize predicted circles that do not appear in the ground-truth. A simple penalty would be to assign an error of 0.5 (i.e., that of a random prediction) to additional circles in the case of Facebook. However, in our experience, our method did not overpredict the number of circles in the case of Facebook: on average, users identified 19 circles, whereas using the Bayesian Information Criterion described in Section \ref{sec:hyper}, our method never predicted $K > 10$. In practice this means that in the case of Facebook, we \emph{always} penalize \emph{all} predictions. Again we note that the process of choosing the number of circles using the BIC is a standard procedure from the literature \citep{airoldi,handcock,volinsky}, whose merit we do not assess in this paper.

\xhdr{Network Modularity}
Although for our algorithm, and other probabilistic baselines, we shall choose the number of communities using the Bayesian Information Criterion as described in Section \ref{sec:hyper}, another standard criterion used to determine the number of communities in a network is the \emph{modularity} \citep{modularity}.

The Bayesian Information Criterion has the advantage that it allows for overlapping communities, whereas the modularity does not (i.e., it assumes all communities are disjoint); it is for this reason that we chose the BIC to choose $\hat{K}$ for our algorithm. On the other hand, the Bayesian Information Criterion can only be computed for \emph{probabilistic} models (i.e., models that associate a likelihood with each prediction), whereas the modularity has no such restriction. For this reason, we shall use the modularity to choose $\hat{K}$ for non-probabilistic baselines.

The \emph{modularity} essentially measures the extent to which clusters in a network have dense internal, but sparse external, connections \citep{newman03fast}. If $e_{ij}$ is the fraction of edges in the network that connect vertices in $C_i$ to vertices in $C_j$, then the modularity is defined as
\begin{equation}
 Q(K) = \sum_{i = 1}^K \left\lbrace e_{ii} - \sum_{j = 1}^K e_{ij} \right\rbrace.
\label{eq:modularity}
\end{equation}
We then choose $\hat{K}$ so that the modularity is maximized.

\subsection{Baselines}
\label{sec:baselines}
We considered a wide number of baseline methods, including those that consider only network structure, those that consider only profile information, and those that consider both.

\xhdr{Mixed Membership Stochastic Block Models} \citep{airoldi}. This method detects communities based only on graph structure; the output is a stochastic vector for each node encoding partial memberships to each community. The optimal number of communities $\hat{K}$ is determined using the Bayesian Information Criterion as described in \eq{eq:barK}. This model is similar to those of \cite{liu} and \cite{chang2009relational}, the latter of which includes the implementation of MMSB that we used. Since we require `hard' memberships for evaluation, we assign a node to a community if its partial membership to that community is positive.

\xhdr{Block-LDA} \citep{blockLDA}. This method is similar MMSB, except that it allows nodes to be augmented with side information in the form of `documents'. For our purposes, we generate `documents' by treating aspects of user profiles as words in a bag-of-words model.

\xhdr{K-means clustering} \citep{MacKay}. Just as MMSB uses only the graph structure, K-means clustering ignores the graph structure and uses only node features (for node features we again use a bag-of-words model). Here, we choose $\hat{K}$ so as to maximize the modularity of $\mathcal C$, as defined in \eq{eq:modularity}. 

\xhdr{Hierarchical Clustering} \citep{hierarchical}. This method builds a hierarchy of clusters. Like K-means, this method form clusters based only on node profiles, but ignores the network.

\xhdr{Link Clustering} \citep{ahn}. Conversely, this method uses network structure, but ignores node features to construct hierarchical communities in networks.

\xhdr{Clique Percolation} \citep{palla}. This method also uses only network structure, and builds communities from the union of small, densely-connected sub-communities.

\xhdr{Low-Rank Embedding} \citep{yoshida}. Uses both graph structure and node similarity information, but does not perform any learning. We adapt an algorithm described by \cite{yoshida}, where node similarities are based on the cosine distance between profile bags-of-words. After our features are embedded into a low-dimensional space, we again use K-means clustering to detect communities, again choosing $\hat{K}$ so as to maximize the modularity.

\xhdr{Multi-Assignment Clustering} \citep{macbd}. Like ours, this method predicts hard assignments to multiple clusters, though it does so without using the network structure.

\begin{table}[t]
  \caption{\mbox{Baselines}}
  \label{tab:baselines}
\small
\begin{center}
\begin{tabular}{|lp{0.11\textwidth}p{0.11\textwidth}p{0.15\textwidth}p{0.15\textwidth}|}
\hline
 Algorithm & network structure? & node/edge features? & overlapping communities? & hard memberships?\\
\hline
 MMSB & Yes & No & Yes & No\\
 Block-LDA & Yes & Yes & Yes & No\\
 K-means & No & Yes & No & Yes\\
 Hierarchical Clustering & No & Yes & Yes & Yes\\
 Link Clustering & Yes & No & No & Yes\\
 Clique Percolation & Yes & No & Yes & Yes\\
 Low-Rank Embedding & Yes & Yes & No & Yes\\
 Multi-Assignment Clustering & No & Yes & Yes & Yes\\
 \textbf{Our algorithm} & \textbf{Yes} & \textbf{Yes} & \textbf{Yes} & \textbf{Yes}\\
\hline
\end{tabular}
\end{center}
\normalsize
\end{table}

The above methods (and our own) are summarized in Table \ref{tab:baselines}.
Of the eight baselines highlighted above we report the three whose overall performance was the best, namely Block-LDA \citep{blockLDA} (which slightly outperformed mixed membership stochastic block models \cite{airoldi}), Low-Rank Embedding \citep{yoshida}, and Multi-Assignment Clustering \citep{macbd}.

\subsection{Performance on Facebook, Google+, and Twitter Data}
\label{sec:performance}
\begin{figure*}[t]
\begin{center}
\hspace{-1.5mm}\includegraphics[width=\textwidth]{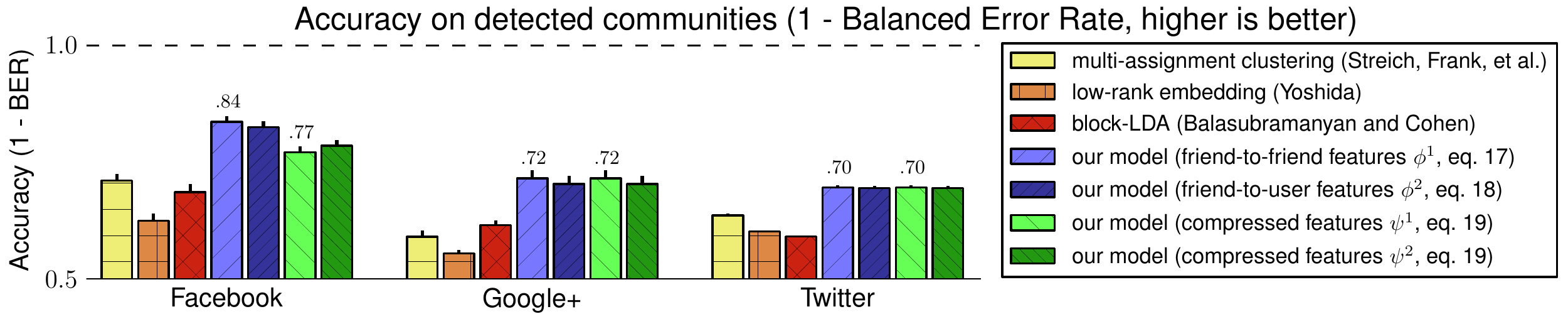}

\vspace{3mm}

\hspace{-1.5mm}\includegraphics[width=\textwidth]{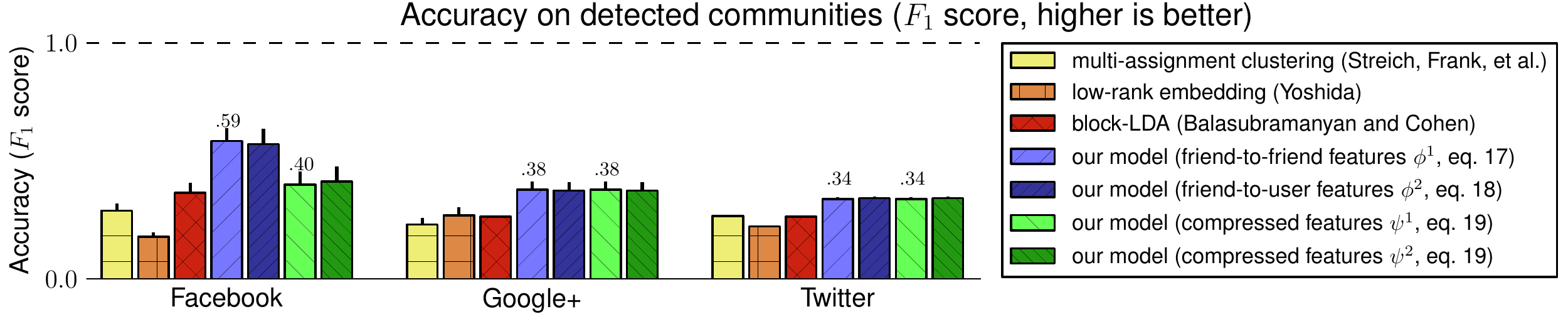}

\end{center}
\caption{Performance on Facebook, Google+, and Twitter, in terms of the Balanced Error Rate (top), and the $F_1$ score (bottom). Higher is better.
Error bars show standard error. The improvement of our best features $\phi^1$ compared to the nearest competitor are significant at the 1\% level or better.
}
\label{fig:results}
\end{figure*}

Figure \ref{fig:results} shows results on our Facebook, Google+, and Twitter data. The largest circles from Google+ were excluded as they exhausted the memory requirements of many of the baseline algorithms. Circles were aligned as described in \eq{eq:delta}, with the number of circles $\hat{K}$ determined as described in Section \ref{sec:learning}. For non-probabilistic baselines, we chose $\hat{K}$ so as to maximize the \emph{modularity}, as described in \eq{eq:modularity}. In terms of absolute performance our best model $\phi^1$ achieves BER scores of 0.84 on Facebook, 0.72 on Google+ and 0.70 on Twitter ($F_1$ scores are 0.59, 0.38, and 0.34, respectively). The lower $F_1$ scores on Google+ and Twitter are explained by the fact that many circles have not been maintained since they were initially created: we achieve high recall (we recover the friends in each circle), but at low precision (we recover additional friends who appeared after the circle was created).

\begin{figure*}[t]
\begin{center}
\parbox[c]{0.095\textwidth}{Facebook:}
\parbox[c]{0.895\textwidth}{\center\includegraphics[angle=-5,trim=0cm 7.5cm 0cm 7.5cm, clip=true, scale=0.18]{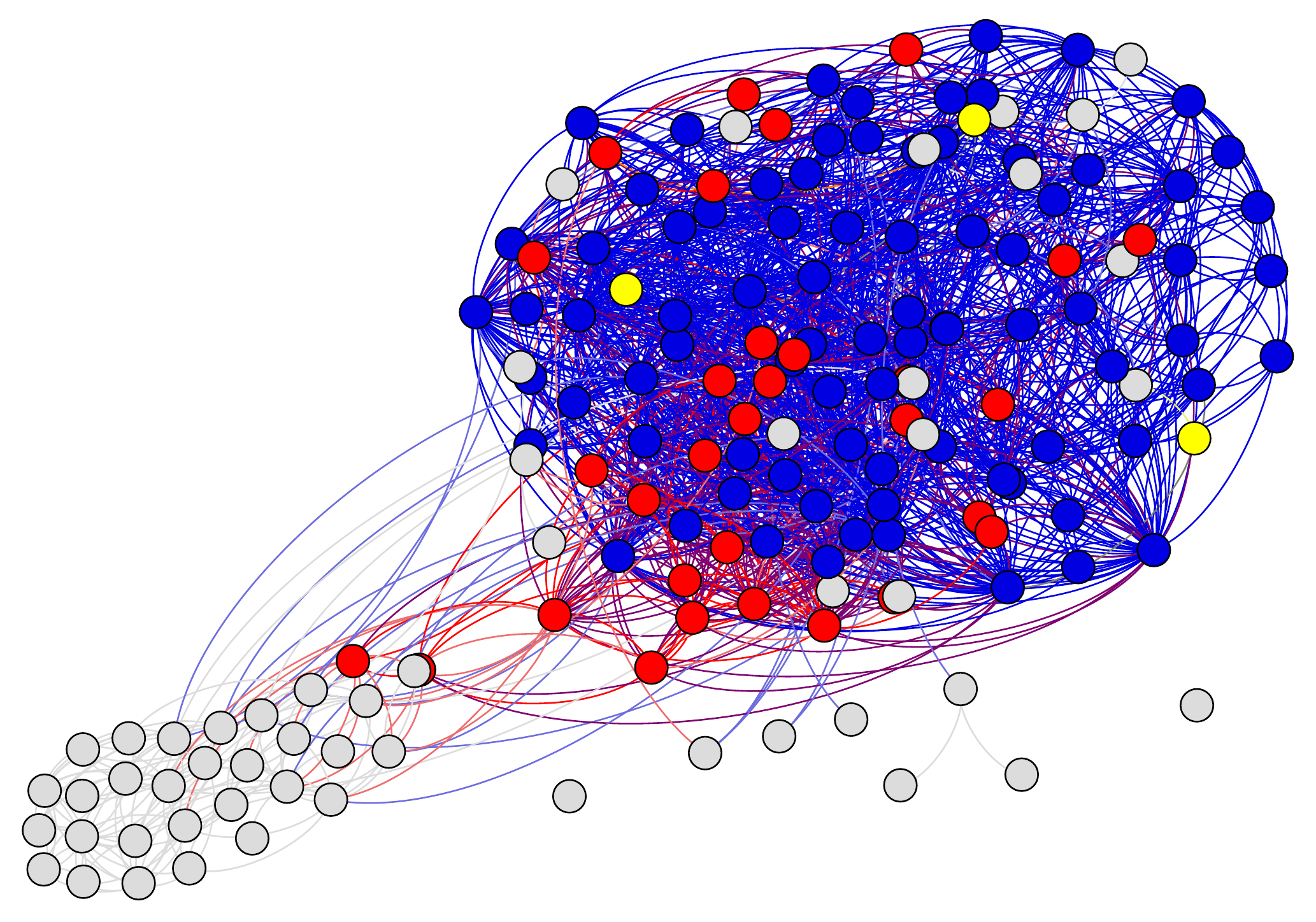}%
\includegraphics[angle=-5,trim=0cm 7.5cm 0cm 7.5cm, clip=true, scale=0.18]{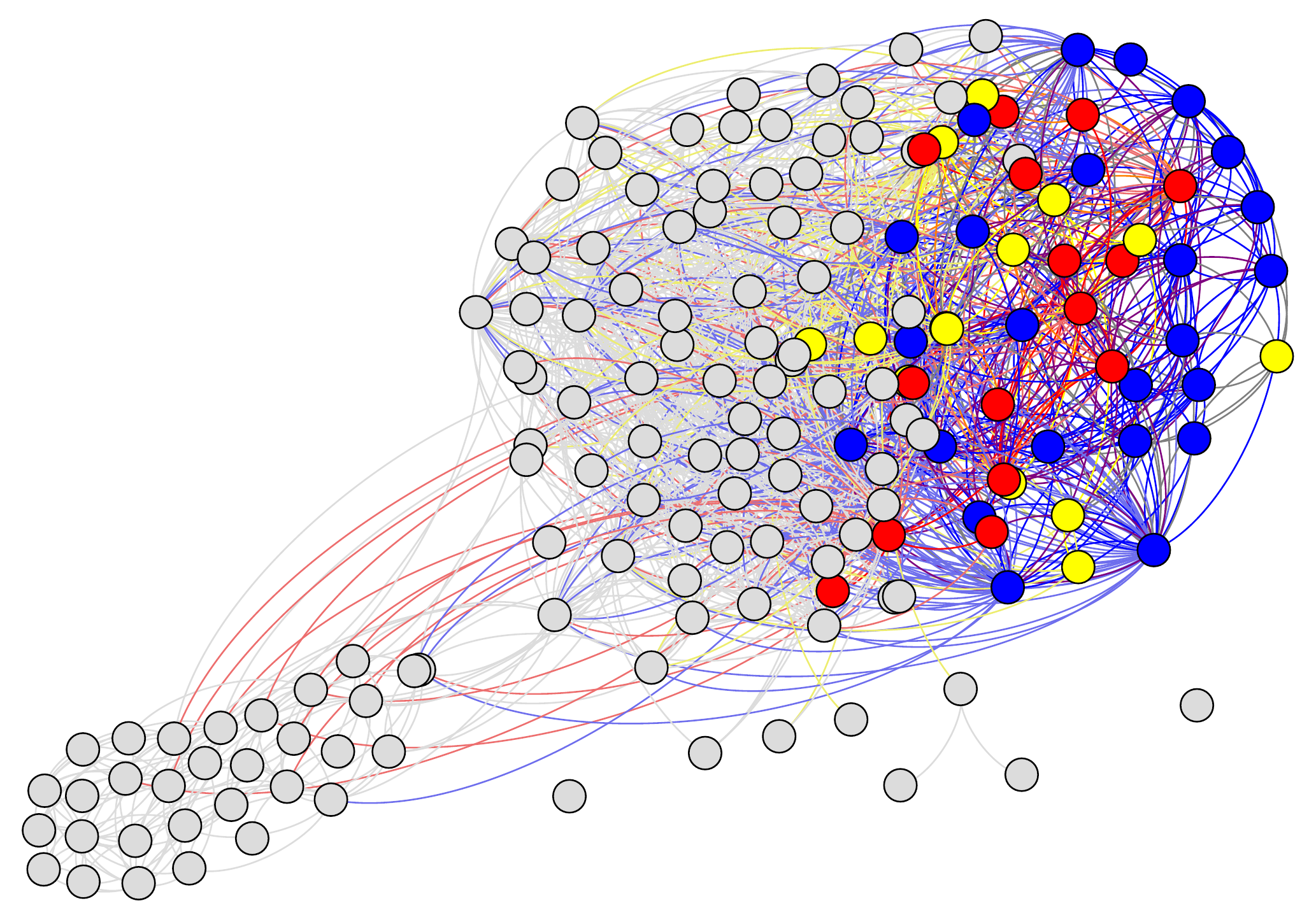}%
\includegraphics[angle=-5,trim=0cm 7.5cm 0cm 7.5cm, clip=true, scale=0.18]{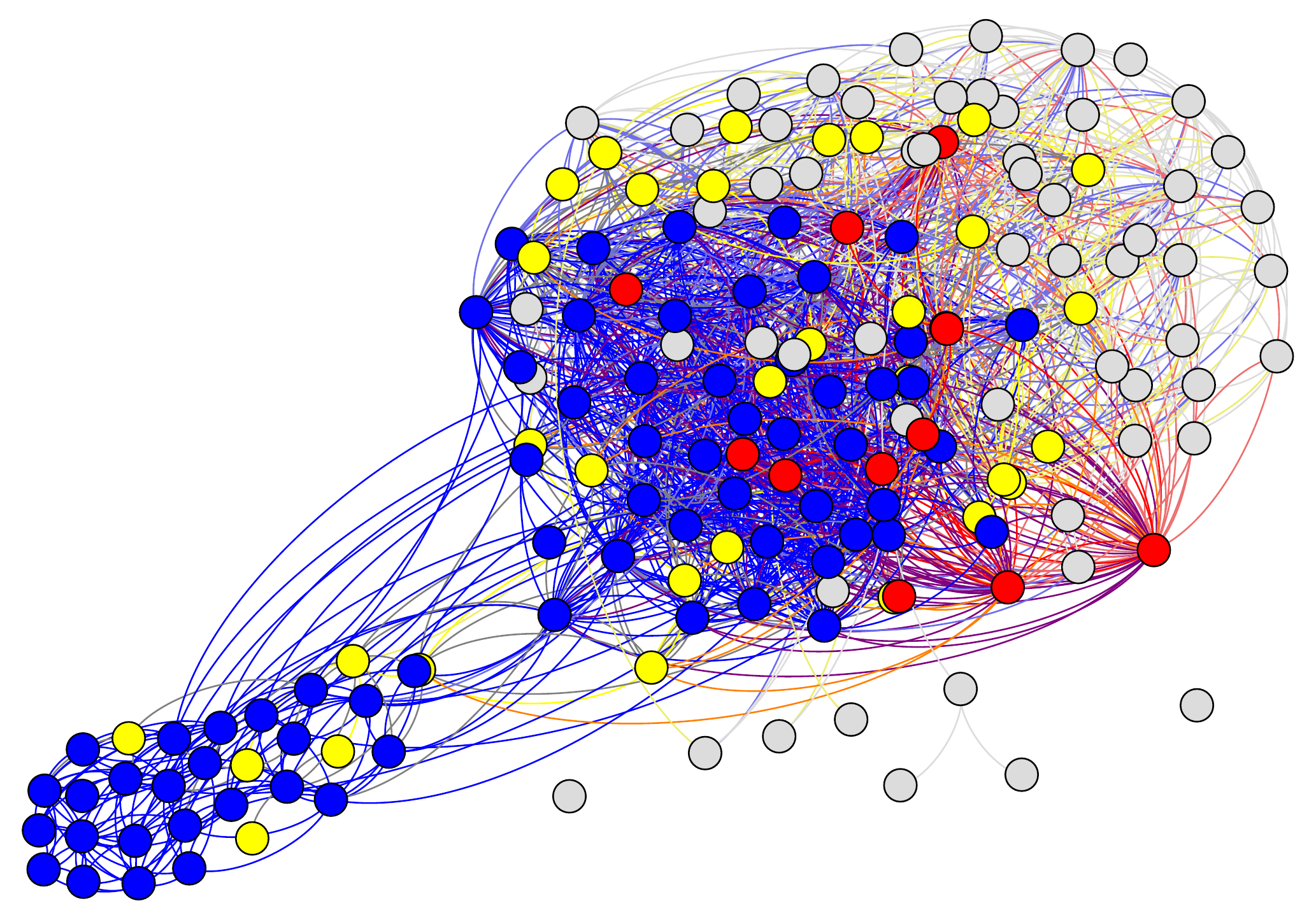}}

\parbox[c]{0.095\textwidth}{Google+:} \parbox[c]{0.895\textwidth}{\center\includegraphics[angle=-5,trim=0cm 10.5cm 9cm 9.5cm, clip=true, scale=0.21]{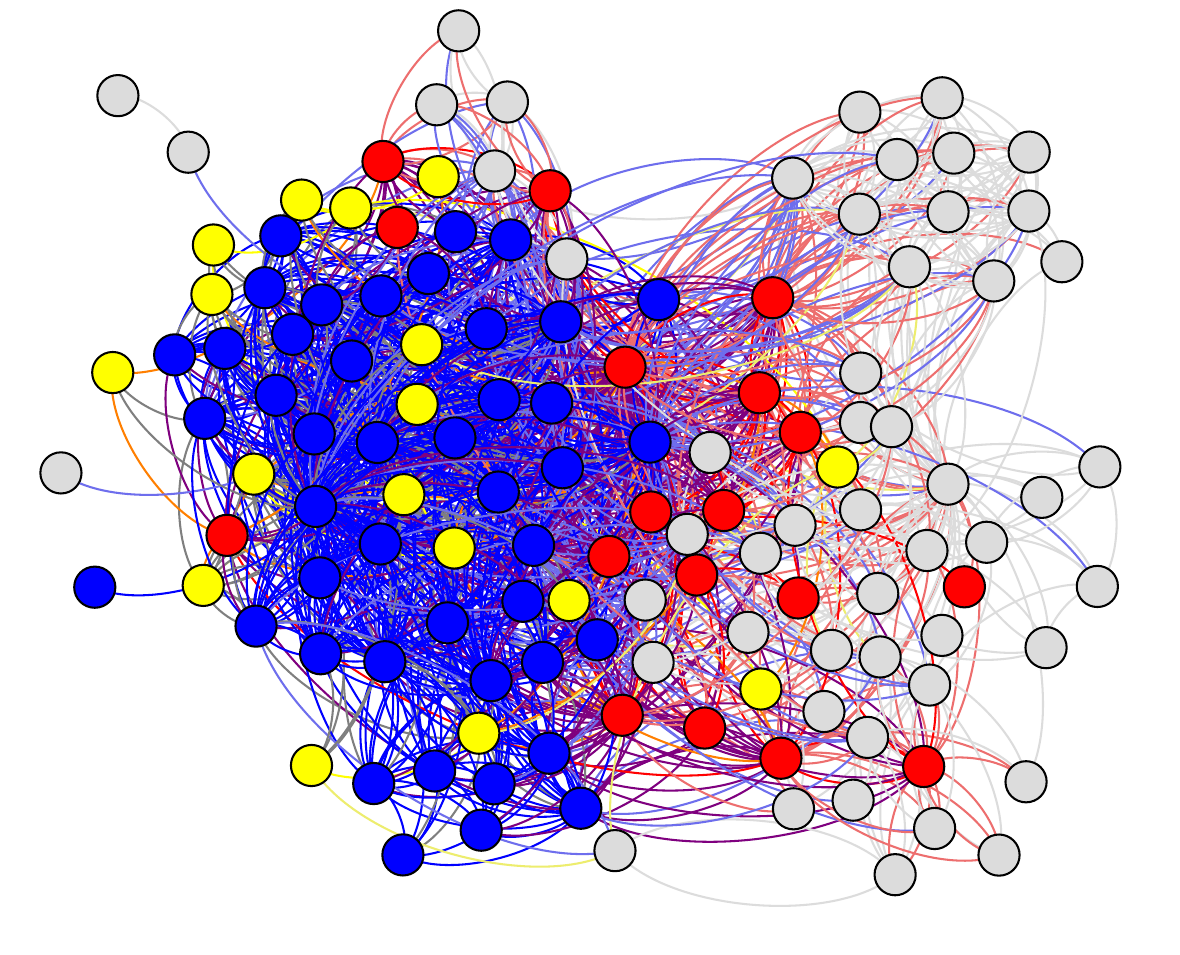}\hspace{5mm}%
\includegraphics[angle=-5,trim=0cm 10.5cm 9cm 9.5cm, clip=true, scale=0.21]{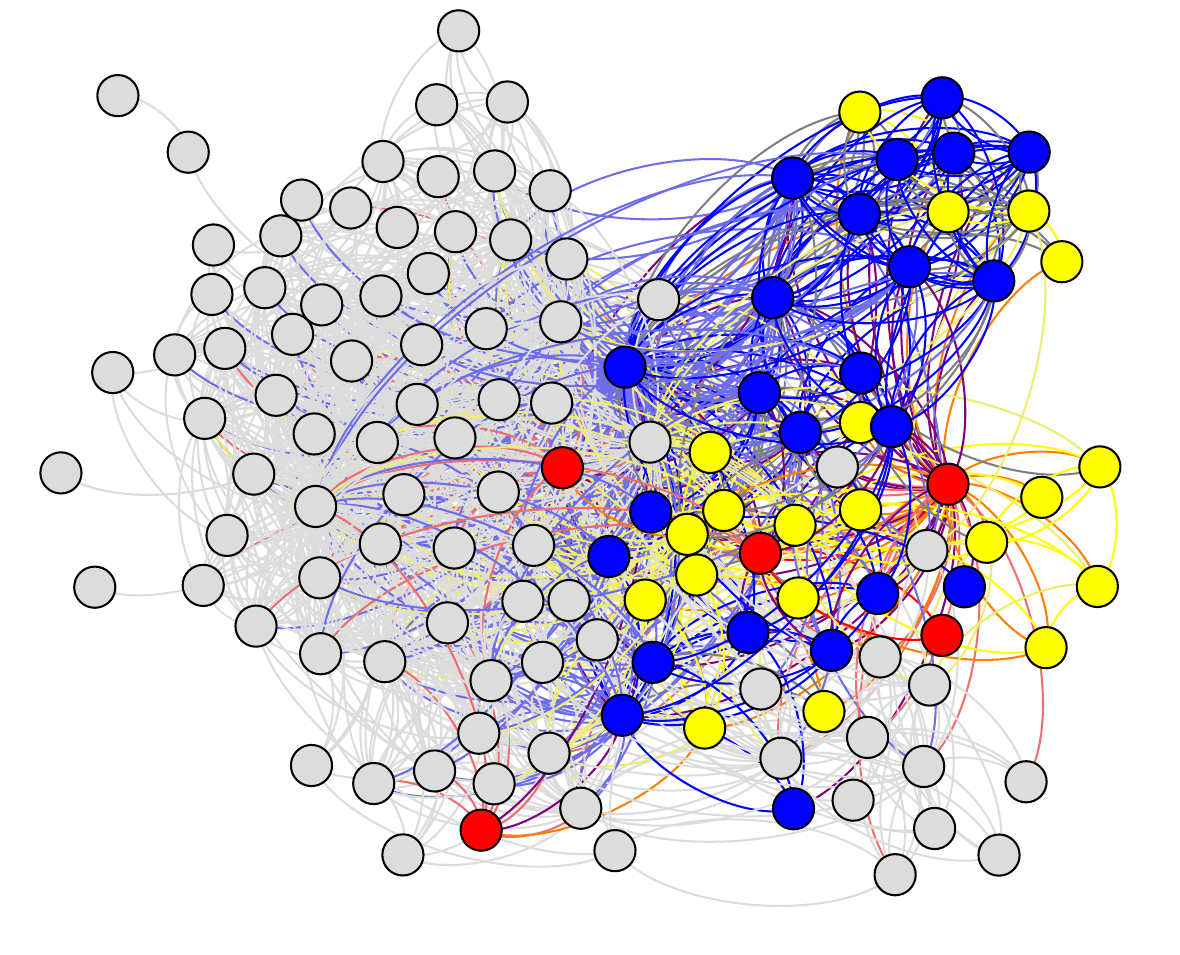}\hspace{5mm}%
\includegraphics[angle=-5,trim=0cm 10.5cm 9cm 9.5cm, clip=true, scale=0.21]{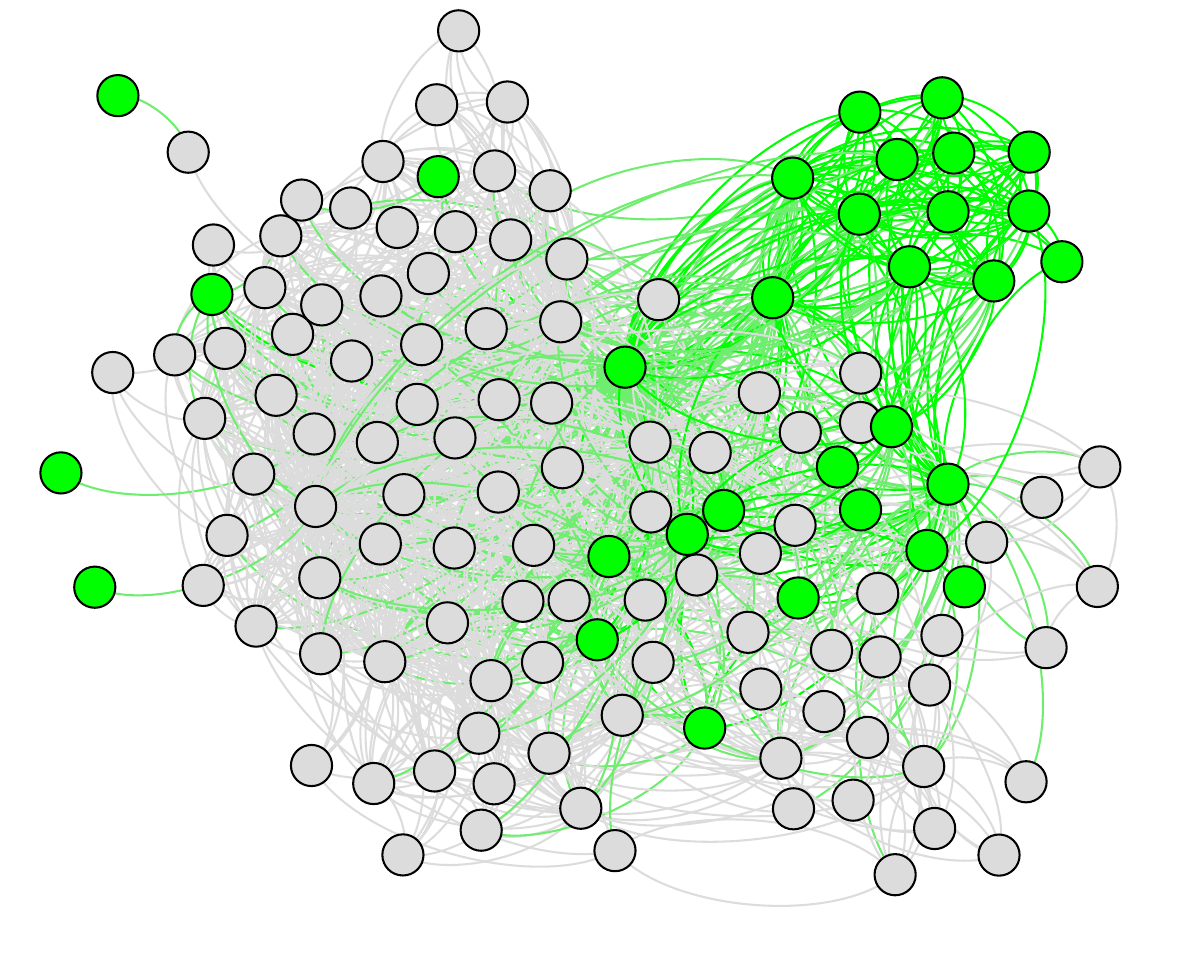}\hspace{5mm}%
\includegraphics[angle=-5,trim=0cm 10.5cm 9cm 9.5cm, clip=true, scale=0.21]{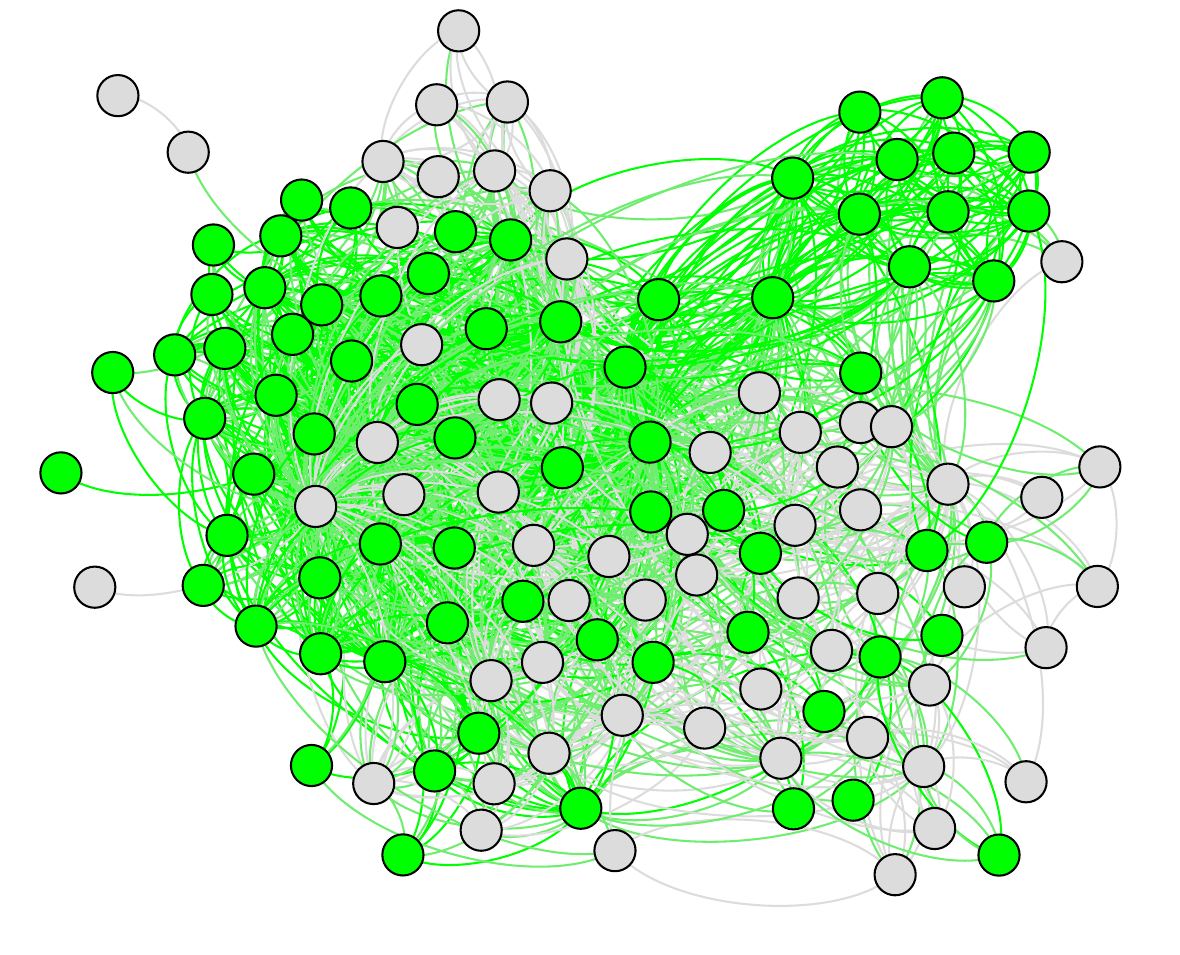}}

\end{center}
\caption{Top: Three detected circles on a small ego-network from Facebook, compared to three ground-truth circles (BER $\simeq 0.81$).
Blue nodes: true positives. Grey: true negatives. Red: false positives. Yellow: false negatives. Our method correctly identifies the largest circle (left), a sub-circle contained within it (center), and a third circle that significantly overlaps with it (right).
Bottom: Four detected circles on ego-networks from Google+ (BER $\simeq 0.73$). Green nodes in the two right networks show additional detected circles, whose accuracy cannot be evaluated as we only observed two circles in the ground-truth.
}
\label{fig:clusters}
\end{figure*}

\begin{figure*}[t]
\begin{center}
 \includegraphics[width=\textwidth]{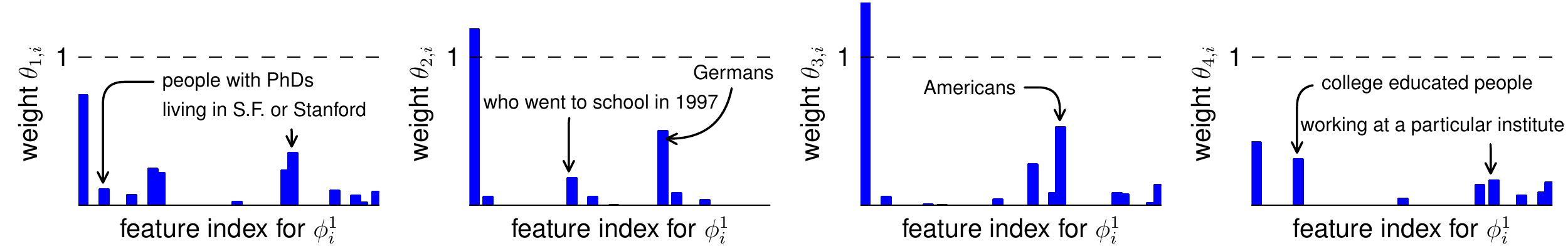}

 \includegraphics[width=\textwidth]{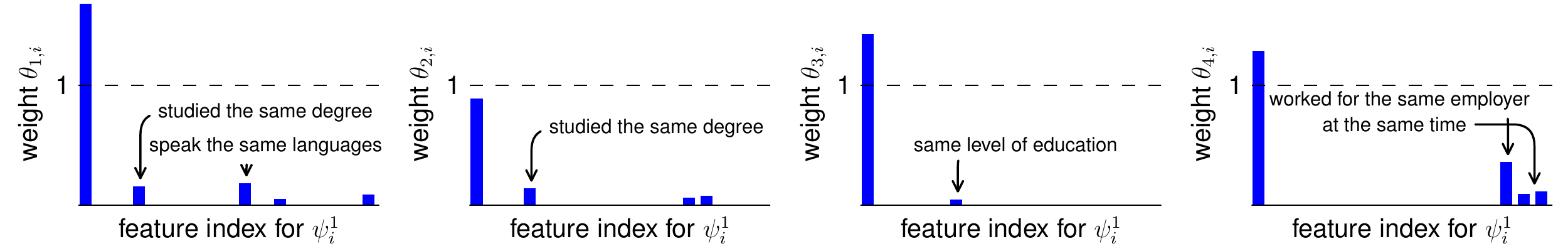}

\end{center}
\caption{Parameter vectors of four communities for a particular Facebook user.
The top four plots show `complete' features $\phi^1$, while the bottom four plots show `compressed' features $\psi^1$ (in both cases, BER $\simeq 0.78$). For example the former features encode the fact that members of a particular community tend to speak German, while the latter features encode the fact that they speak the same language. (Personally identifiable annotations have been suppressed.)}
\label{fig:weights}
\end{figure*}

Comparing our method to baselines we notice that we outperform all baselines on all datasets by a statistically significant margin. Compared to the nearest competitors, our best performing features $\phi^1$ improve on the BER by 43\% on Facebook, 26\% on Google+, and 16\% on Twitter (improvements in terms of the $F_1$ score are similar). Regarding the performance of the baseline methods, we note that good performance seems to depend critically on predicting {\emph{hard}} memberships to \emph{multiple} circles, using a combination of \emph{node and edge} information; none of the baselines from Table \ref{tab:baselines} exhibit precisely this combination, a shortcoming our model addresses.

Both of the features we propose (friend-to-friend features $\phi^1$ and friend-to-user features $\phi^2$) perform similarly, revealing that both schemes ultimately encode similar information, which is not surprising, since users and their friends have similar profiles. Using the `compressed' features $\psi^1$ and $\psi^2$ does not significantly impact performance, which is promising since they have far lower dimension than the full features; what this reveals is that it is sufficient to model \emph{categories} of attributes that users have in common (e.g.~same school, same town), rather than the attribute values themselves.

We found that all algorithms perform significantly better on Facebook than on Google+ or Twitter. There are a few explanations: Firstly, our Facebook data is \emph{complete}, in the sense that survey participants manually labeled \emph{every} circle in their ego-networks, whereas in other datasets we only observe publicly-visible circles, which may not be up-to-date. Secondly, the 26 profile categories available from Facebook are more informative than the 6 categories from Google+, or the tweet-based profiles we built from Twitter. A more basic difference lies in the nature of the networks themselves: edges in Facebook encode \emph{mutual} ties, whereas edges in Google+ and Twitter encode follower relationships, which changes the role that circles serve \citep{twitterLists}. The latter two points explain why algorithms that use either edge or profile information in isolation are unlikely to perform well on this data.

\subsection{Qualitative Analysis}
\label{sec:qualitative}
Next we examine the output of our model in greater detail.
Figure \ref{fig:clusters} shows results of our unsupervised method on example ego-networks from Facebook and Google+. Different colors indicate \mbox{true-,} false- positives and negatives. Our method is correctly able to identify overlapping circles as well as sub-circles (circles within circles). 

Figure \ref{fig:weights} shows parameter vectors learned for four circles for a particular Facebook user. Positive weights indicate properties that users in a particular circle have in common. Notice how the model naturally learns the social dimensions that lead to a social circle. Moreover, the first parameter that corresponds to a constant feature `1' has the highest weight; this reveals that membership to the same community provides the strongest signal that edges will form, while profile data provides a weaker (but still relevant) signal.

\subsection{Circle Maintenance}
\label{sec:results_maintenance}

Next we examine the problem of adding new users to already-defined ego-networks, in which complete circles have already been provided. For evaluation, we suppress a single user $u$ from a user's ego-network, and learn the model parameters $\hat{\Theta}$ that best fit $G \setminus \lbrace u \rbrace$ and $\mathcal C \setminus \lbrace u \rbrace$. Our goal is then to recover the set of communities to which the node $u$ belongs, as described in Section \ref{sec:newnode}. Again we report the Balanced Error Rate and $F_1$ score between the ground-truth and the predicted set of community memberships for $u$. We use all of each users' circles for training, up to a maximum of fifteen circles. This experiment is repeated for 10 random choices of the user $u$ for each ego-network in our dataset.

As a baseline we compare the performance of our algorithm to that of a fully-supervised Support Vector Machine (SVM) model. For each community $C_k$, we train a binary classifier that discriminates members from non-members based on their node features. Binary classifications are then made for each community independently.

\begin{figure*}[t]
\begin{center}

\hspace{-1.5mm}\includegraphics[width=\textwidth]{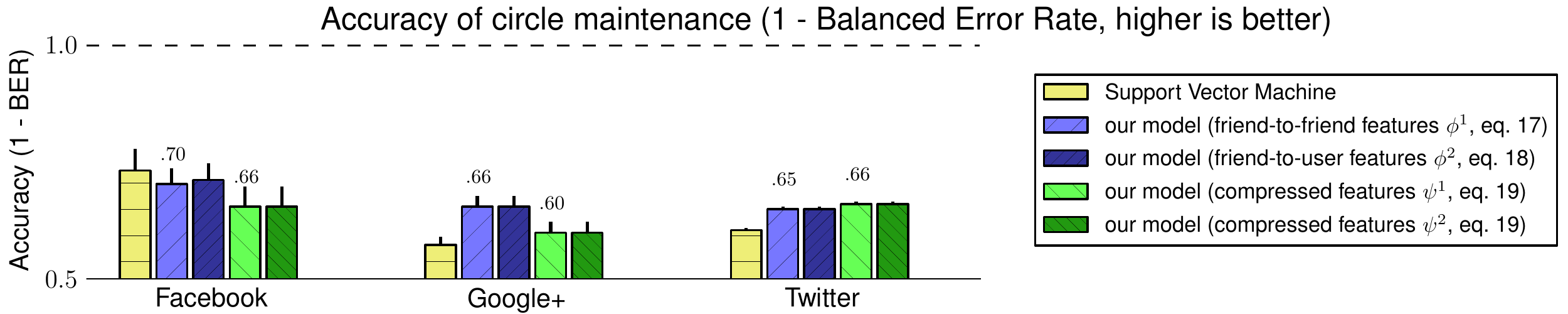}
\vspace{3mm}

\hspace{-1.5mm}\includegraphics[width=\textwidth]{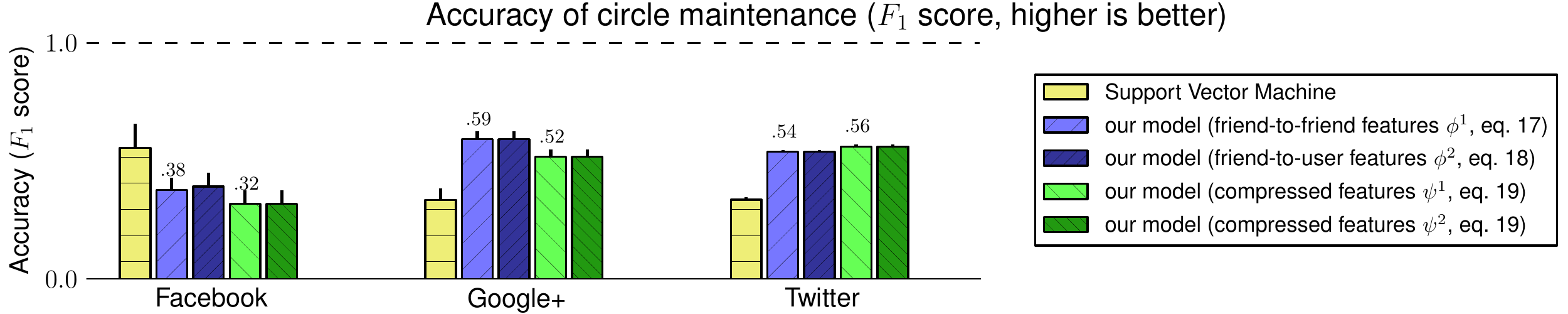}

\end{center}
\caption{Accuracy of assigning a new node to already-existing circles. Although a fully-supervised Support Vector Machine gives accurate results on Facebook (where node features are highly informative), our model yields far better results on Google+ and Twitter data.}
\label{fig:newnode}
\end{figure*}

Performance on this task is shown in Figure \ref{fig:newnode}. On Facebook, Google+, and Twitter our best performing features $\phi^1$ achieve Balanced Error Rates of 0.30, 0.34, and 0.34 (respectively), and $F_1$ scores of 0.38, 0.59, and 0.54. The SVM model achieves better accuracy when rich node features are available (which is the case for Facebook), though it fails to make use of edge information, and does not account for interdependencies between circles. This proves critical in the case of Google+ and Twitter, where node information alone proves uninformative.

\subsection{Semi-Supervised Circle Prediction}
\label{sec:results_semisupervised}

Our next task is to identify circles using a form of weak supervision provided by the user, in the form of \emph{seed nodes} as described in Section \ref{sec:seednodes}. In this setting, the user provides $S$ seed nodes for each of $K$ circles that they wish to identify. For evaluation, we select the $K$ circles to be identified and the $S$ seed nodes uniformly at random.

Without seed nodes (as in our initial experiments), the circles that are automatically identified by our algorithm may be quite different from those identified once seed nodes are added. Similarly, there may be many circles containing the same seed nodes, meaning that different solutions may be chosen for different values of $S$. Thus it is difficult to compare the loss of \eq{eq:delta} with and without seed nodes. To address this, we modify the matching objective of \eq{eq:delta} so that the $K$ circles randomly selected for seeding must be the same as those matched when evaluating the loss. Thus the loss is always evaluated on the same $K$ circles for every number of seed nodes $S \in \lbrace 0 \ldots 10 \rbrace$. Note also that for each value of $K$, performance is only evaluated on those ego-networks with at least $K$ ground-truth circles.

\begin{figure*}[t]

\begin{tabular}{m{0.1\textwidth}m{0.68\textwidth}m{0.06\textwidth}}
           & \hspace{2.5mm}\parbox{0.17\textwidth}{\begin{center}\scriptsize{friend-to-friend\\ features $\phi^1$}\end{center}}\parbox{0.17\textwidth}{\begin{center}\scriptsize{friend-to-user\\ features $\phi^2$}\end{center}}\parbox{0.17\textwidth}{\begin{center}\scriptsize{compressed\\ features $\psi^1$}\end{center}}\parbox{0.17\textwidth}{\begin{center}\scriptsize{compressed\\ features $\psi^2$}\end{center}}\\
 Facebook: & \includegraphics[width=0.17\textwidth]{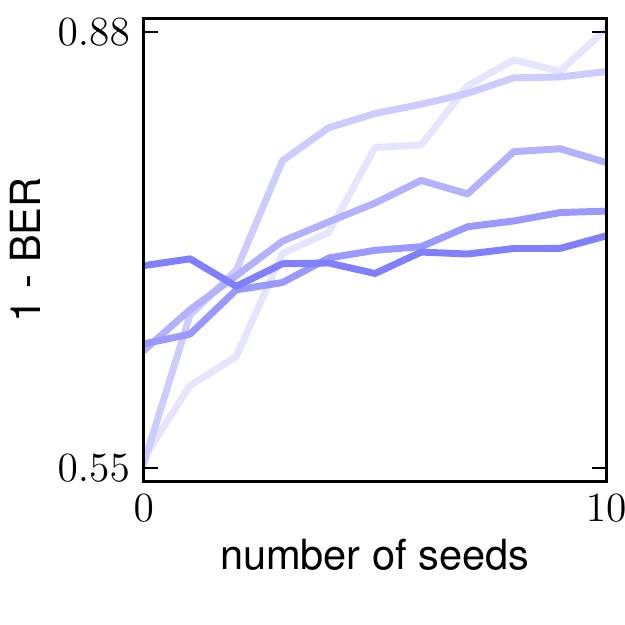}\includegraphics[width=0.17\textwidth]{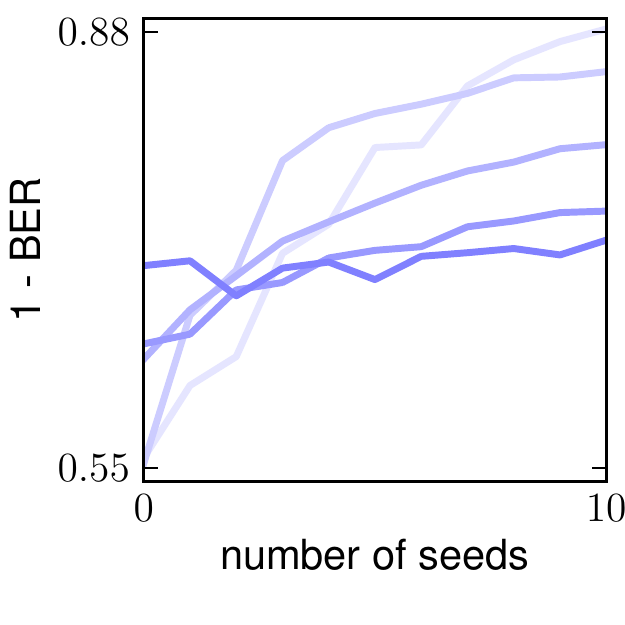}\includegraphics[width=0.17\textwidth]{facebook_edge_ber}\includegraphics[width=0.17\textwidth]{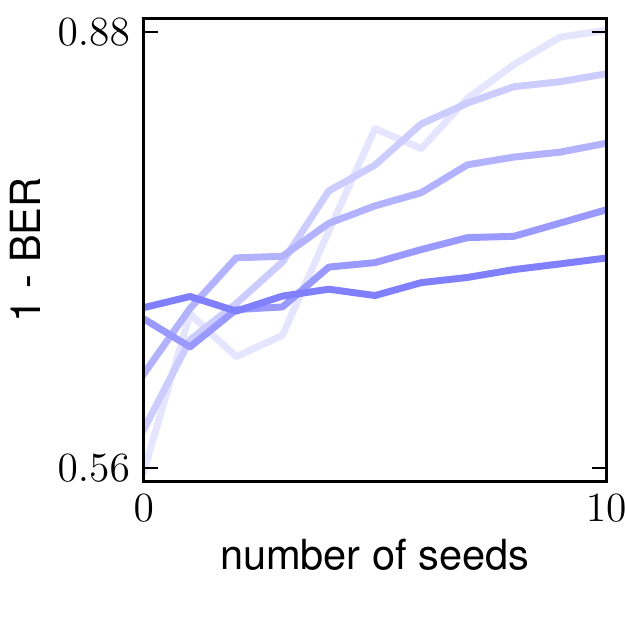}\\
 Google+: & \includegraphics[width=0.17\textwidth]{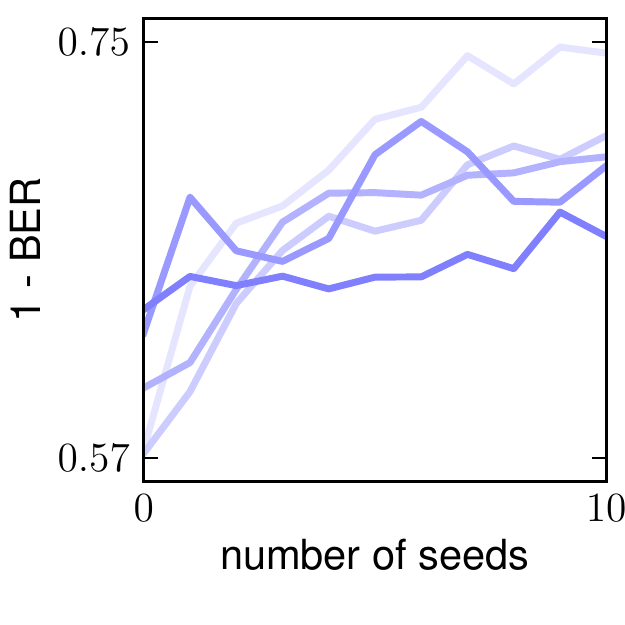}\includegraphics[width=0.17\textwidth]{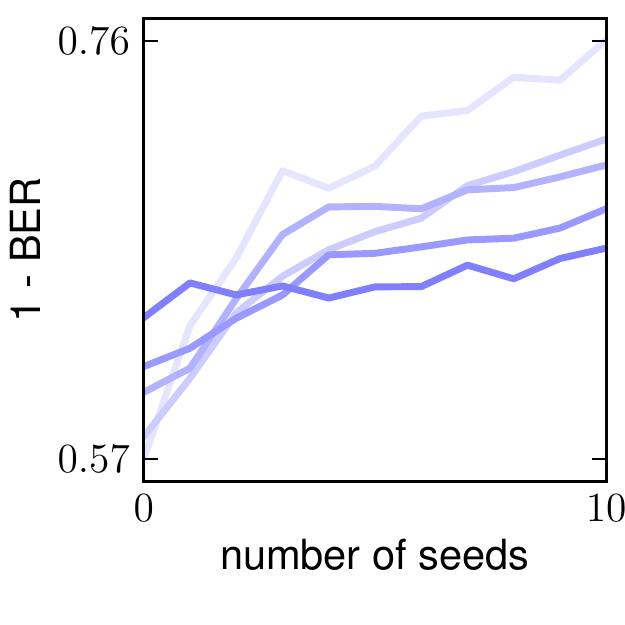}\includegraphics[width=0.17\textwidth]{gplus_edge_ber}\includegraphics[width=0.17\textwidth]{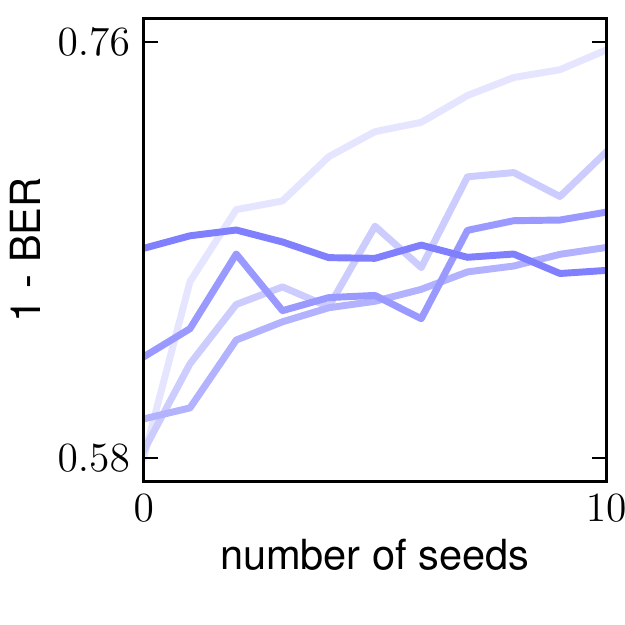} & \vspace{-8mm}\hspace{-6mm}\includegraphics[width=0.125\textwidth]{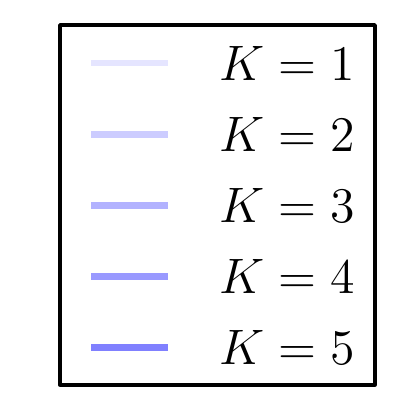}\\
 Twitter: & \includegraphics[width=0.17\textwidth]{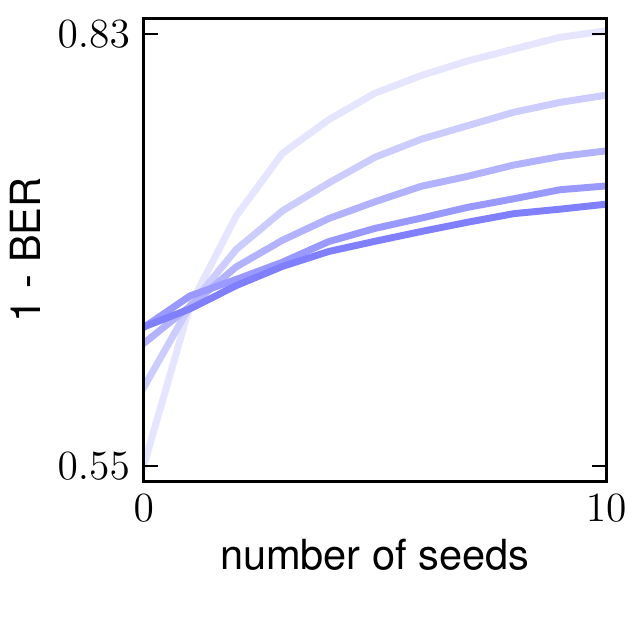}\includegraphics[width=0.17\textwidth]{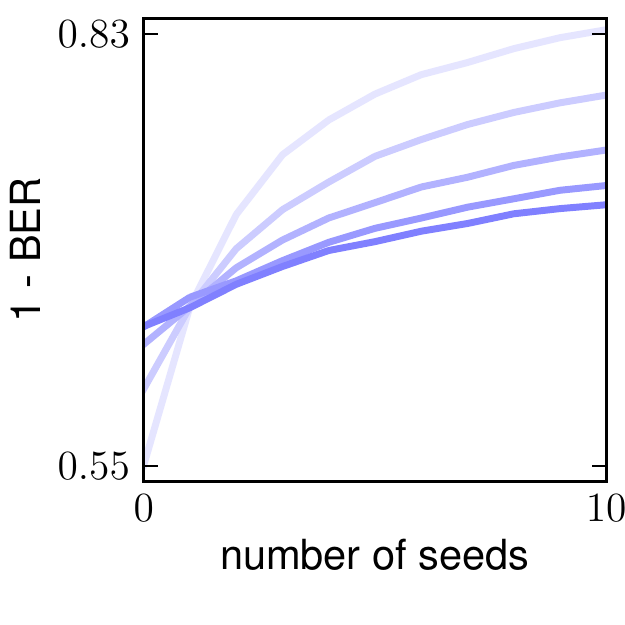}\includegraphics[width=0.17\textwidth]{twitter_edge_ber}\includegraphics[width=0.17\textwidth]{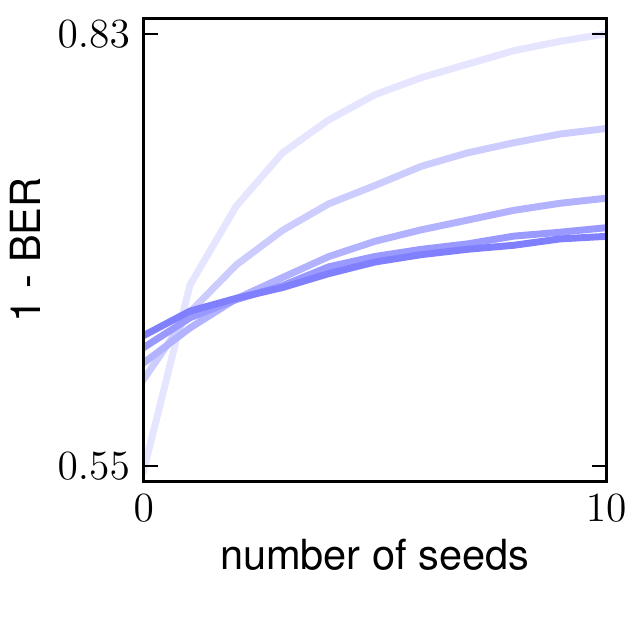}\\
\end{tabular}

\caption{Number of seeds versus accuracy (1 - Balanced Error Rate) for different numbers of circles $K$. For each of the $K$ circles being identified, the user provides the same number of seeds. Although providing additional seeds is generally beneficial to performance for all $K$, the benefit is most pronounced when the number of circles to be identified is small. Results in terms of the $F_1$ score are qualitatively similar and are omitted for brevity.}
\label{fig:seednodes}
\end{figure*}

Figure \ref{fig:seednodes} shows the performance of our algorithm for different numbers of seed nodes $S \in \lbrace 0 \ldots 10 \rbrace$ and different numbers of circles $K \in \lbrace 1 \ldots 5 \rbrace$. The same results in terms of the $F_1$ score are qualitatively similar and are omitted for brevity. We find that for all values of $K$, adding seed nodes increases the accuracy significantly, though the effect is most pronounced when the number of circles that the user wishes to identify is small.

Curiously, we find that while larger values of $K$ lead to better prediction when there are no seeds, the opposite is true when there are many seeds. The former behavior may be explained by the simple fact that larger values of $K$ are better able to fit the data, though the latter behavior is more enigmatic. Pleasingly, assuming that a user wishes to identify only a small number of circles at a time, then they can do so with very few seeds: for small $K$, most of the benefit is gained once only two or three seeds are provided.

\subsection{Scalability Analysis}
\label{sec:results_scalability}

\begin{figure*}[t]
\begin{center}

\includegraphics[width=\textwidth]{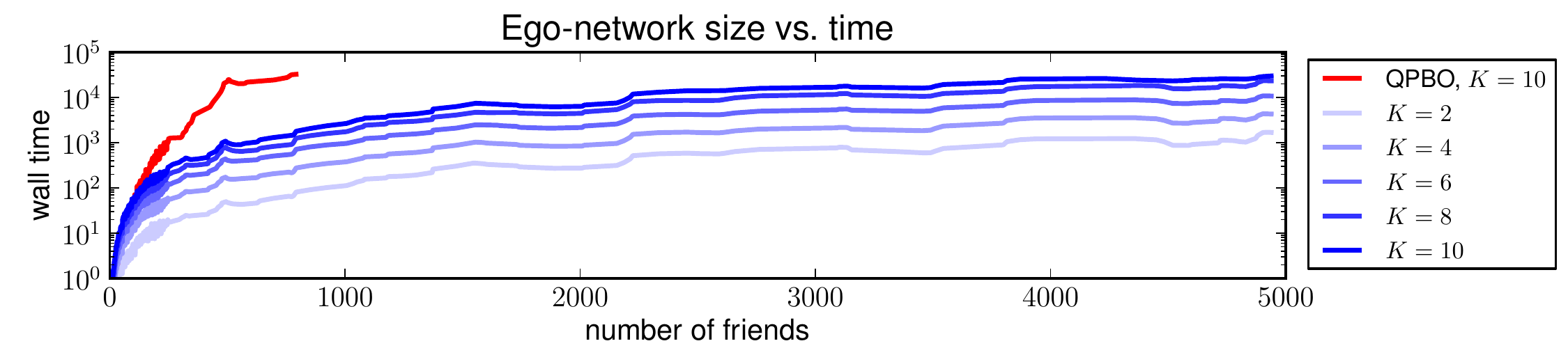}

%

\end{center}
\caption{Running time of our Markov Chain Monte Carlo (MCMC) algorithm for different ego-network sizes and different values of $K$ (the number of circles to be detected). For comparison, our previously described inference algorithm (based on QPBO \citep{rother07}) is shown for $K = 10$.}
\label{fig:scalability_walltime}
\end{figure*}

Figure \ref{fig:scalability_walltime} examines how our algorithm scales with the size of an ego-network. Here we use the Markov-Chain Monte-Carlo (MCMC) version of our algorithm described in Section \ref{sec:fast}. Figure \ref{fig:scalability_walltime} shows the total time taken to predict different numbers of circles in differently sized ego-networks. Since the performance of our MCMC algorithm is a function of the number of circles $K$ and the feature dimensionality $F$, we fix the feature dimensionality at $F = 10$ for all ego-networks, using the ten most common features that appear in each ego-network using the `friend-to-friend' features $\phi^1$.

For comparison, Figure \ref{fig:scalability_walltime} shows the running time of inference using QPBO as described in Section \ref{sec:learning}. Although the two algorithms are competitive for up to a few hundred nodes, the QPBO algorithm becomes intractable for networks of around 1000 nodes, since it requires us to optimize a probability distribution defined on \emph{complete} graphs (in practice, in order to apply the QPBO algorithm in the previous experiments, we did not construct complete graphs, but rather included only those edges whose influence on the likelihood was maximal).

\begin{figure*}[t]
\begin{center}

\includegraphics[width=\textwidth]{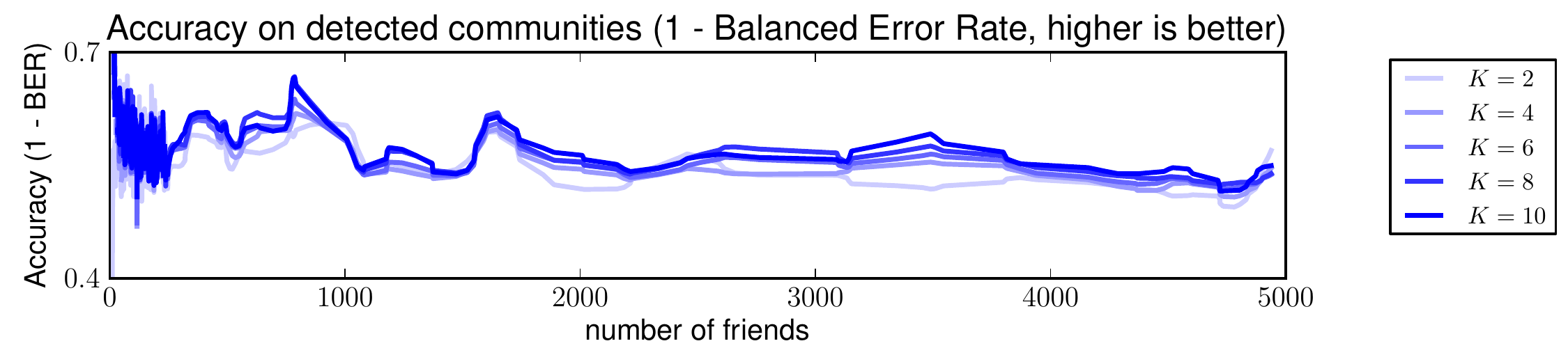}

\includegraphics[width=\textwidth]{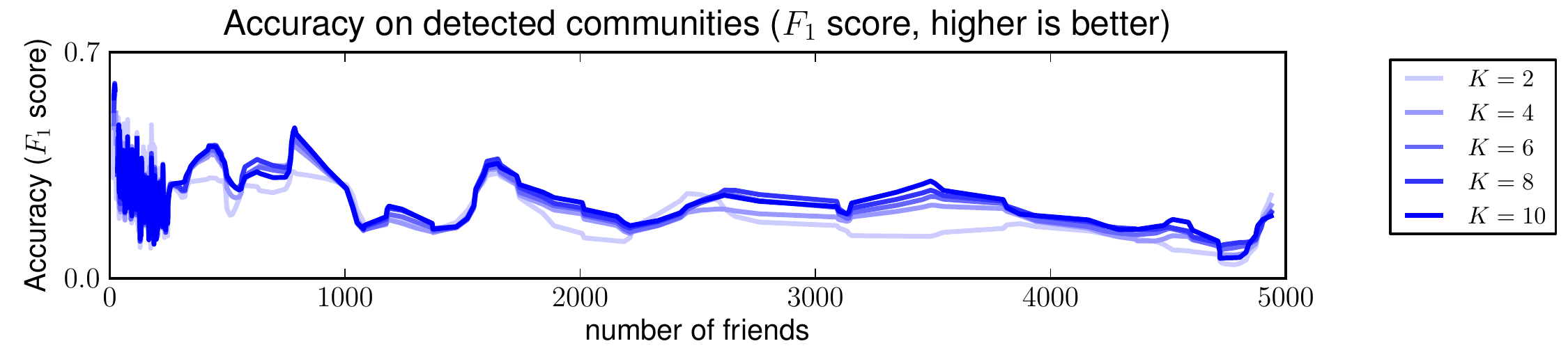}

\end{center}
\caption{Accuracy of our Markov Chain Monte Carlo (MCMC) algorithm, in terms of the Balanced Error Rate (top), and the $F_1$ score (bottom).}
\label{fig:scalability_performance}
\end{figure*}

Although this version of the algorithm is not particularly efficient for small networks (identifying $K = 10$ circles on an ego network with 1000 nodes requires around one hour), it has the advantage that it is easily able to scale to the largest ego-networks that are ever encountered. For very large networks, the algorithm is able to take advantage of the fact that many nodes with the same features and community memberships can be `collapsed', so that the running time increases only modestly between 2500 and 5000 node ego-networks.

Figure \ref{fig:scalability_performance} shows the accuracy of our MCMC algorithm in terms of the Balanced Error Rate and $F_1$ score. We note that the best performance of our algorithm is obtained on reasonably small ego-networks, though in practice small networks account for the vast majority of our data. Note that the results for any particular value of $K$ are slightly worse than those reported in Figure \ref{fig:results}, since we are not selecting $K$ using the BIC described in Section \ref{sec:hyper}. Although performance clearly degrades for large ego-networks, it remains an open question whether this is due to the difficulty of optimization on large networks, or simply due to the fact that our model assumptions become increasingly violated as large networks become less `community-like'.

\section{Discussion and Future Work}

We have modeled circle detection as a problem that can be solved independently for each user. In practice this assumption is advantageous, as it allows us to deal with several small problems independently, using sophisticated models that could not easily scale to networks with millions of nodes. However, it is possible that circles could be more accurately predicted by exploiting relationships between the circles of multiple users. For example, if a user has a `Stanford' circle in their ego-network, it is highly likely that users belonging to that circle will \emph{also} have Stanford circles within their own ego-networks. Alternately, if a Stanford \emph{community} could be detected across the entire Facebook, Google+, or Twitter network, then a user's `Stanford' circle might simply be the intersection of their ego-network with that community. Although studying such models is an appealing avenue for future work, it is unfortunately not possible using our data, where we do not have access to complete network information.

Although we developed algorithms that scale to the largest ego-networks that we encountered, we find that the best performance occurs on ego-networks with up to a few hundred nodes, but degrades significantly for networks with more than 1000. It remains to be seen whether this is a shortcoming of our algorithm (due to the fact that optimization is more difficult for large networks), or whether the assumptions of our model simply break down at large scales. Our fundamental assumption that circles will be made up of close-knit groups of friends with common properties seems like a better fit to networks with at most a few hundred nodes.

We also found that performance on even the largest Facebook networks (i.e., over 1000 friends) was better than that obtained on small networks from Google+ and Twitter. This suggests that it is not merely the size of the networks that causes our model assumptions to become violated, but rather the very nature of the networks themselves (in addition to the differences in the ground-truth already mentioned). Naturally, a circle containing members of the same squash team (as we find on Facebook) is fundamentally different from a circle containing presidential candidates (as we find on Google+). It remains to design a circle detection algorithm that is tailored for networks with asymmetric following relationships.

\section{Conclusion}

`Circles' allow us to organize the overwhelming volumes of data generated by our personal social networks, though they are laborious to construct manually. We have designed an algorithm to automatically detect circles in ego-networks, which we evaluated on a dataset of 1,143 ego-networks and 5,541 ground-truth circles obtained from Facebook, Google+, and Twitter. We find in such data circles that are disjoint, overlapping, and hierarchically nested, and design our model with such behavior in mind. Our model is unsupervised, but can also make use of weakly-labeled data that may be available in practice. Experiments reveal that social circles can be accurately detected using a combination of both network and profile information.

\section*{Acknowledgements} This research has been supported in part by NSF
IIS-1016909,                                                                                            
CNS-1010921,                                                                                            
CAREER IIS-1149837,                                                                                     
IIS-1159679,                                                                                            
AFRL FA8650-10-C-7058,                                                                  
DARPA XDATA,                                                                                                  
DARPA GRAPHS,                                                                                                 
Albert Yu \& Mary Bechmann Foundation,  
Boeing,                                                                                                                                 
Allyes,                                                                                                                                 
Samsung,                                                                                                                                
Intel,                                  
Alfred P. Sloan Fellowship and                                  
the Microsoft Faculty Fellowship.                       

The U.S. Government is authorized to reproduce and distribute reprints for Governmental purposes notwithstanding any copyright annotation thereon. [Disclaimer] The views and conclusions contained herein are those of the authors and should not be interpreted as necessarily representing the official policies or endorsements, either expressed or implied, of IARPA, AFRL, or the U.S. Government.

\bibliographystyle{plainnat}
\bibliography{arxiv}

\begin{thebibliography}{60}
\providecommand{\natexlab}[1]{#1}
\providecommand{\url}[1]{\texttt{#1}}
\expandafter\ifx\csname urlstyle\endcsname\relax
  \providecommand{\doi}[1]{doi: #1}\else
  \providecommand{\doi}{doi: \begingroup \urlstyle{rm}\Url}\fi

\bibitem[Agarwal et~al.(2008)Agarwal, Chen, Elango, Motgi, Park, Ramakrishnan,
  Roy, and Zachariah]{agarwal}
D.~Agarwal, B.-C. Chen, P.~Elango, N.~Motgi, S.-T. Park, R.~Ramakrishnan,
  S.~Roy, and J.~Zachariah.
\newblock Online models for content optimization.
\newblock In \emph{Neural Information Processing Systems}, 2008.

\bibitem[Ahn et~al.(2010)Ahn, Bagrow, and Lehmann]{ahn}
Y.-Y. Ahn, J.~Bagrow, and S.~Lehmann.
\newblock Link communities reveal multiscale complexity in networks.
\newblock \emph{Nature}, 2010.

\bibitem[Airoldi et~al.(2008)Airoldi, Blei, Fienberg, and Xing]{airoldi}
E.~Airoldi, D.~Blei, S.~Fienberg, and E.~Xing.
\newblock Mixed membership stochastic blockmodels.
\newblock \emph{Journal of Machine Learning Research}, 2008.

\bibitem[Andersen and Lang(2006)]{andersen06seed}
R.~Andersen and K.~Lang.
\newblock Communities from seed sets.
\newblock In \emph{WWW}, 2006.

\bibitem[Backstrom et~al.(2006)Backstrom, Huttenlocher, Kleinberg, and
  Lan]{lars06groups}
L.~Backstrom, D.~Huttenlocher, J.~Kleinberg, and X.~Lan.
\newblock Group formation in large social networks: membership, growth, and
  evolution.
\newblock In \emph{Proceedings of the 12th ACM SIGKDD International Conference
  on Knowledge Discovery and Data Mining}, 2006.

\bibitem[Balasubramanyan and Cohen(2011)]{blockLDA}
R.~Balasubramanyan and W.~Cohen.
\newblock Block-{LDA}: Jointly modeling entity-annotated text and entity-entity
  links.
\newblock In \emph{SIAM International Conference on Data Mining}, 2011.

\bibitem[Boros and Hammer(2002)]{boros}
E.~Boros and P.~Hammer.
\newblock Pseudo-boolean optimization.
\newblock \emph{Discrete Applied Mathematics}, 2002.

\bibitem[Chang and Blei(2009)]{chang2009relational}
J.~Chang and D.~Blei.
\newblock Relational topic models for document networks.
\newblock In \emph{International Conference on Artificial Intelligence and
  Statistics}, 2009.

\bibitem[Chang et~al.(2009)Chang, Boyd-Graber, and Blei]{nubbi}
J.~Chang, J.~Boyd-Graber, and D.~Blei.
\newblock Connections between the lines: augmenting social networks with text.
\newblock In \emph{Knowledge Discovery and Data Mining}, 2009.

\bibitem[Chen and Karger(2006)]{harr}
H.~Chen and R.~Karger.
\newblock Less is more: Probabilistic models for retrieving fewer relevant
  documents.
\newblock In \emph{Special Interest Group on Information Retrieval}, 2006.

\bibitem[Chen and Lin(2006)]{chen06}
Y.~Chen and C.~Lin.
\newblock \emph{Combining {SVM}s with various feature selection strategies}.
\newblock Springer, 2006.

\bibitem[El-Arini et~al.(2009)El-Arini, Veda, Shahaf, and Guestrin]{khalid}
K.~El-Arini, G.~Veda, D.~Shahaf, and C.~Guestrin.
\newblock Turning down the noise in the blogosphere.
\newblock In \emph{Knowledge Discovery and Data Mining}, 2009.

\bibitem[Feld(1981)]{feld86focused}
Scott~L. Feld.
\newblock The focused organization of social ties.
\newblock \emph{American Journal of Sociology}, 1981.

\bibitem[Frank et~al.(2012)Frank, Streich, Basin, and Buhmann]{macbd}
Mario Frank, Andreas~P. Streich, David Basin, and Joachim~M. Buhmann.
\newblock Multi-assignment clustering for {B}oolean data.
\newblock \emph{Journal of Machine Learning Research}, 2012.

\bibitem[Gregory(2010{\natexlab{a}})]{Gregory10LabelOverlap}
Steve Gregory.
\newblock Finding overlapping communities in networks by label propagation.
\newblock \emph{New Journal of Physics}, 2010{\natexlab{a}}.

\bibitem[Gregory(2010{\natexlab{b}})]{gregory}
Steve Gregory.
\newblock Fuzzy overlapping communities in networks.
\newblock \emph{CoRR}, abs/1010.1523, 2010{\natexlab{b}}.

\bibitem[Hammer et~al.(1984)Hammer, Hansen, and Simeone]{hammer84}
P.~Hammer, P.~Hansen, and B.~Simeone.
\newblock {R}oof duality, complementation and persistency in quadratic 0-1
  optimization.
\newblock \emph{Mathematical Programming}, 1984.

\bibitem[Handcock et~al.(2007{\natexlab{a}})Handcock, Raftery, and
  Tantrum]{handcock}
M.~Handcock, A.~Raftery, and J.~Tantrum.
\newblock Model-based clustering for social networks.
\newblock \emph{Journal of the Royal Statistical Society Series A},
  2007{\natexlab{a}}.

\bibitem[Handcock et~al.(2007{\natexlab{b}})Handcock, Raftery, and
  Tantrum]{handcock07}
Mark~S. Handcock, Adrian~E. Raftery, and Jeremy~M. Tantrum.
\newblock Model-based clustering for social networks.
\newblock \emph{Journal of the Royal Statistical Society}, 2007{\natexlab{b}}.

\bibitem[Hastings(2006)]{hastings06_inference}
M.~B. Hastings.
\newblock Community detection as an inference problem.
\newblock \emph{Physical Review E}, 2006.

\bibitem[Haussler(1999)]{haussler99convolution}
David Haussler.
\newblock Convolution kernels on discrete structures.
\newblock Technical report, University of California at Santa Cruz, 1999.

\bibitem[Hoff et~al.(2002)Hoff, Raftery, and Handcock]{Hoff02latentspace}
Peter~D. Hoff, Adrian~E. Raftery, and Mark~S. Handcock.
\newblock Latent space approaches to social network analysis.
\newblock \emph{Journal of the American Statistical Association}, 2002.

\bibitem[Johnson(1967)]{hierarchical}
S.~Johnson.
\newblock Hierarchical clustering schemes.
\newblock \emph{Psychometrika}, 1967.

\bibitem[Kim et~al.(2010)Kim, Jo, Moon, and Oh]{lists3}
D.~Kim, Y.~Jo, L.-C. Moon, and A.~Oh.
\newblock Analysis of twitter lists as a potential source for discovering
  latent characteristics of users.
\newblock In \emph{CHI}, 2010.

\bibitem[Kohli and Torr(2005)]{kohli05}
P.~Kohli and P.~Torr.
\newblock Efficiently solving dynamic {Markov} random fields using graph cuts.
\newblock In \emph{International Conference on Computer Vision}, 2005.

\bibitem[Kolmogorov and Rother(2007)]{kolmo07}
Vladimir Kolmogorov and Carsten Rother.
\newblock Minimizing nonsubmodular functions with graph cuts-a review.
\newblock \emph{IEEE Transactions on Pattern Analysis and Machine
  Intelligence}, 2007.

\bibitem[Krivitsky et~al.(2009)Krivitsky, Handcock, Raftery, and Hoff]{hoff09}
P.~Krivitsky, M.~Handcock, A.~Raftery, and P.~Hoff.
\newblock Representing degree distributions, clustering, and homophily in
  social networks with latent cluster random effects models.
\newblock \emph{Social Networks}, 2009.

\bibitem[Lancichinetti and Fortunato(2009{\natexlab{a}})]{LF09_TR}
A.~Lancichinetti and S.~Fortunato.
\newblock Community detection algorithms: a comparative analysis.
\newblock arXiv:0908.1062, 2009{\natexlab{a}}.

\bibitem[Lancichinetti and
  Fortunato(2009{\natexlab{b}})]{lancichinetti09ovlpbenchmark}
Andrea Lancichinetti and Santo Fortunato.
\newblock Benchmarks for testing community detection algorithms on directed and
  weighted graphs with overlapping communities.
\newblock \emph{Physical Review E}, 2009{\natexlab{b}}.

\bibitem[Lancichinetti et~al.(2009)Lancichinetti, Fortunato, and
  Kertesz]{Lancichinetti09overlapping}
Andrea Lancichinetti, Santo Fortunato, and Janos Kertesz.
\newblock Detecting the overlapping and hierarchical community structure in
  complex networks.
\newblock \emph{New Journal of Physics}, 2009.

\bibitem[Lazarsfeld and Merton(1954)]{homophily1}
P.~Lazarsfeld and R.~Merton.
\newblock Friendship as a social process: A substantive and methodological
  analysis.
\newblock In \emph{Freedom and Control in Modern Society}. 1954.

\bibitem[Leskovec et~al.(2010)Leskovec, Lang, and Mahoney]{jure10community}
J.~Leskovec, K.~Lang, and M.~Mahoney.
\newblock Empirical comparison of algorithms for network community detection.
\newblock In \emph{WWW}, 2010.

\bibitem[Liu et~al.(2009)Liu, Niculescu-Mizil, and Gryc]{liu}
Y.~Liu, A.~Niculescu-Mizil, and W.~Gryc.
\newblock Topic-link {LDA}: joint models of topic and author community.
\newblock In \emph{International Conference on Machine Learning}, 2009.

\bibitem[MacKay(2003)]{MacKay}
D.~MacKay.
\newblock \emph{Information Theory, Inference and Learning Algorithms}.
\newblock Cambrdige University Press, 2003.

\bibitem[McAuley and Leskovec(2012)]{nips2012}
J.~McAuley and J.~Leskovec.
\newblock Learning to discover social circles in ego networks.
\newblock In \emph{Neural Information Processing Systems}, 2012.

\bibitem[McPherson(1983)]{blau}
M.~McPherson.
\newblock An ecology of affiliation.
\newblock \emph{American Sociological Review}, 1983.

\bibitem[McPherson et~al.(2001)McPherson, Smith-Lovin, and
  Cook]{mcpherson01homophily}
Miller McPherson, Lynn Smith-Lovin, and James~M. Cook.
\newblock Birds of a feather: Homophily in social networks.
\newblock \emph{Annual Review of Sociology}, 2001.

\bibitem[Menon and Elkan(2011)]{menon}
A.~Menon and C.~Elkan.
\newblock Link prediction via matrix factorization.
\newblock In \emph{European Conference on Machine Learning and Principles and
  Practice of Knowledge Discovery in Databases}, 2011.

\bibitem[Menon and Elkan(2010)]{menonICDM}
Aditya Menon and Charles Elkan.
\newblock A log-linear model with latent features for dyadic prediction.
\newblock In \emph{International Conference on Data Mining}, 2010.

\bibitem[Mislove et~al.(2010)Mislove, Viswanath, Gummadi, and
  Druschel]{mislove2010}
A.~Mislove, B.~Viswanath, K.~Gummadi, and P.~Druschel.
\newblock You are who you know: Inferring user profiles in online social
  networks.
\newblock In \emph{International Conference on Web Search and Data Mining},
  2010.

\bibitem[Nasirifard and Hayes(2011)]{tadvise}
P.~Nasirifard and C.~Hayes.
\newblock Tadvise: A twitter assistant based on twitter lists.
\newblock In \emph{SocInfo}, 2011.

\bibitem[Newman(2006)]{modularity}
M.~Newman.
\newblock Modularity and community structure in networks.
\newblock \emph{Proceedings of the National Academy of Sciences}, 2006.

\bibitem[Newman(2003)]{newman03fast}
M.~E.~J. Newman.
\newblock Fast algorithm for detecting community structure in networks.
\newblock \emph{Physical Review E}, 2003.

\bibitem[Newman(2004)]{newman2004_detect}
M.~E.~J. Newman.
\newblock Detecting community structure in networks.
\newblock \emph{The European Physical Journal B}, 2004.

\bibitem[Newman and Barkema(1999)]{Newman99MonteCarlo}
M.~E.~J. Newman and G.~T. Barkema.
\newblock \emph{Monte Carlo Methods in Statistical Physics}.
\newblock Oxford University Press, 1999.

\bibitem[Nocedal(1980)]{lbfgs}
Jorge Nocedal.
\newblock Updating quasi-{Newton} matrices with limited storage.
\newblock \emph{Mathematics of Computation}, 1980.

\bibitem[Palla et~al.(2005)Palla, Derenyi, Farkas, and Vicsek]{palla}
G.~Palla, I.~Derenyi, I.~Farkas, and T.~Vicsek.
\newblock Uncovering the overlapping community structure of complex networks in
  nature and society.
\newblock \emph{Nature}, 2005.

\bibitem[Porter et~al.(2009)Porter, Onnela, and Mucha]{POM09_TR}
M.~A. Porter, J.-P. Onnela, and P.~J. Mucha.
\newblock Communities in networks.
\newblock February 2009.

\bibitem[Ravasz and Barab\'{a}si(2003)]{ravasz03_hierarchical}
E.~Ravasz and A.-L. Barab\'{a}si.
\newblock Hierarchical organization in complex networks.
\newblock \emph{Physical Review E}, 2003.

\bibitem[Rother et~al.(2007)Rother, Kolmogorov, Lempitsky, and
  Szummer]{rother07}
C.~Rother, V.~Kolmogorov, V.~Lempitsky, and M.~Szummer.
\newblock Optimizing binary {MRFs} via extended roof duality.
\newblock In \emph{Computer Vision and Pattern Recognition}, 2007.

\bibitem[Schaeffer(2007)]{Schaeffer07_survey}
S.E. Schaeffer.
\newblock Graph clustering.
\newblock \emph{Computer Science Review}, 2007.

\bibitem[Simmel(1964)]{simmel64affiliations}
Georg Simmel.
\newblock \emph{Conflict and the web of group affiliations}.
\newblock Simon and Schuster, 1964.

\bibitem[Ugander et~al.(2011)Ugander, Karrer, Backstrom, and Marlow]{nfriends}
J.~Ugander, B.~Karrer, L.~Backstrom, and C.~Marlow.
\newblock The anatomy of the {Facebook} social graph.
\newblock preprint, 2011.

\bibitem[Vishwanathan and Smola(2002)]{treekernels}
S.~V.~N. Vishwanathan and Alexander~J. Smola.
\newblock Fast kernels for string and tree matching.
\newblock In \emph{Neural Information Processing Systems}, 2002.

\bibitem[Volinsky and Raftery(2000)]{volinsky}
C.~Volinsky and A.~Raftery.
\newblock Bayesian information criterion for censored survival models.
\newblock \emph{Biometrics}, 2000.

\bibitem[Vu et~al.(2011)Vu, Asuncion, Hunter, and Smyth]{dvu}
D.~Vu, A.~Asuncion, D.~Hunter, and P.~Smyth.
\newblock Dynamic egocentric models for citation networks.
\newblock In \emph{International Conference on Machine Learning}, 2011.

\bibitem[Wu et~al.(2011)Wu, Hofman, Mason, and Watts]{twitterLists}
S.~Wu, J.~Hofman, W.~Mason, and D.~Watts.
\newblock Who says what to whom on twitter.
\newblock In \emph{WWW}, 2011.

\bibitem[Yang and Leskovec(2012)]{jaewon2012}
J.~Yang and J.~Leskovec.
\newblock Community-affiliation graph model for overlapping community
  detection.
\newblock In \emph{International Conference on Data Mining}, 2012.

\bibitem[Yoshida(2010)]{yoshida}
T.~Yoshida.
\newblock Toward finding hidden communities based on user profiles.
\newblock In \emph{ICDM Workshops}, 2010.

\bibitem[Zhao(2011)]{twitterLists2}
J.~Zhao.
\newblock Examining the evolution of networks based on lists in twitter.
\newblock In \emph{International Conference on Internet Multimedia System
  Architectures and Applications}, 2011.

\end{thebibliography}

\end{document}